\documentclass{article}

\usepackage{arxiv}
\usepackage[utf8]{inputenc} 
\usepackage[T1]{fontenc}    
\usepackage{hyperref}       
\usepackage{url}            
\usepackage{booktabs}       
\usepackage{amsfonts}       
\usepackage{nicefrac}       
\usepackage{microtype}      
\usepackage{lipsum}		
\usepackage{graphicx}
\usepackage{doi}

\usepackage{amsmath}
\usepackage{amsthm}
\usepackage{graphics}
\usepackage{marvosym}  
\usepackage{subfigure}
\usepackage{url}
\usepackage{amssymb}
\usepackage[ruled,linesnumbered]{algorithm2e}
\usepackage{color,array}
\DeclareMathOperator*{\argmax}{argmax}
\title{Variational Auto-encoder Based Solutions to Interactive Dynamic Influence Diagrams}

\author{  
  \href{https://orcid.org/0000-0001-5715-2855}{Yinghui Pan~\includegraphics[scale=0.06]{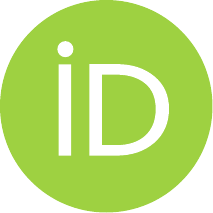}
}\\  
  National Engineering Laboratory for Big Data System Computing Technology \\  
  Shenzhen University\\
  Shenzhen, China \\  
  \texttt{panyinghui@szu.edu.cn} \and 
  \href{https://orcid.org/0000-0003-1515-6449}{Biyang Ma~\includegraphics[scale=0.06]{orcid.pdf} }\\ 
  School of Computer Science\\
  Minnan Normal University\\
  Zhangzhou, China \\ 
  \texttt{mby@mnnu.edu.cn} \and 
  Hanyi Zhang \\  
  National Engineering Laboratory for Big Data System Computing Technology \\  
  Shenzhen University\\
  Shenzhen, China \\  
  \and 
  \href{https://orcid.org/0000-0002-5246-403X}{Yifeng Zeng~\includegraphics[scale=0.06]{orcid.pdf}} \\  
  Department of Computer and Information Sciences \\  
  Northumbria University, Newcastle, UK \\  
  \texttt{yifeng.zeng@northumbria.ac.uk}  
}

\renewcommand{\shorttitle}{Variational Auto-encoder Based Solutions to Interactive Dynamic Influence Diagrams}

\hypersetup{
pdftitle={Variational Auto-encoder Based Solutions to Interactive Dynamic Influence Diagrams},
pdfsubject={q-bio.NC, q-bio.QM},
pdfauthor={Yinghui Pan, Biyang Ma, Hanyi Zhang, Yifeng Zeng},
pdfkeywords={Decision-making, Multi-agent systerm, Variational autoencoder},
}

\begin{document}
\maketitle

\begin{abstract}
	Addressing multiagent decision problems in AI, especially those involving collaborative or competitive agents acting concurrently in a partially observable and stochastic environment, remains a formidable challenge. While Interactive Dynamic Influence Diagrams~(I-DIDs) have offered a promising decision framework for such problems, they encounter limitations when the subject agent encounters unknown behaviors exhibited by other agents that are not explicitly modeled within the I-DID. This can lead to sub-optimal responses from the subject agent. In this paper, we propose a novel data-driven approach that utilizes an encoder-decoder architecture, particularly a variational autoencoder, to enhance I-DID solutions. By integrating a perplexity-based tree loss function into the optimization algorithm of the variational autoencoder, coupled with the advantages of Zig-Zag One-Hot encoding and decoding, we generate potential behaviors of other agents within the I-DID that are more likely to contain their true behaviors, even from limited interactions. This new approach enables the subject agent to respond more appropriately to unknown behaviors, thus improving its decision quality. 
We empirically demonstrate the effectiveness of the proposed approach in two well-established problem domains, highlighting its potential for handling multi-agent decision problems with unknown behaviors. This work is the first time of using neural networks based approaches to deal with the I-DID challenge in agent planning and learning problems.
\end{abstract}


\section{Introduction}
\label{sec:introduction}
Interactions between intelligent agents operating in a shared and uncertain environment amplify the overall system's uncertainty, posing formidable challenges for efficient modeling and decision-making in multi-agent systems. Consequently, the actions of these agents mutually influence each other, necessitating the comprehensive consideration of both environmental dynamics and potential action sequences of other agents in order to make optimal decisions~\cite{ref_36_8365805}. This issue is commonly referred to as multi-agent sequential decision making under uncertainty~\cite{ref_25_10.1007/11527862_33}.

Over the years, various decision models have been proposed to tackle multi-agent sequential decision making~(MSDM) problems, including decentralized partially observable Markov decision processes~(POMDPs) \cite{ref_33_Seuken07:Formal}, interactive POMDPs \cite{ref_32_Gmytrasiewicz05:Framework:JAIR}, and interactive dynamic influence diagrams~(I-DIDs) \cite{ref_31_Doshi09:Graphical}. Among these, I-DIDs stand out for their ability to model both collaborative and competitive agents, as well as their computational advantages \cite{ref_29_Zeng12:Exploiting, ref_30_DoshiGD20}. Furthermore, their probabilistic graphical model structure makes them inherently explainable in reasoning about the behaviors of other agents.

To solve I-DIDs, knowledge-based approaches have been explored to provide the subject agent with more information about other agents \cite{ref_23_Pan2015,ref_24_Pan2015I,ref_22_Zeng2012}. However, these approaches may be limited by environmental complexity, incomplete data, and uncertainty, resulting in inaccurate or limited information. One approach to address this is by increasing the number of candidate models for other agents, but this can lead to a prohibitively large model set, increasing computational complexity \cite{ref_34_Chen2015,ref_26_Zeng2016Approximating}. As an alternative, data-driven methods have been proposed that leverage historical behavioral data of other agents to learn their potential decision models, enhancing the adaptability and interpret-ability of the decision models \cite{ref_6_Pan2021TowardDS}.

However, data-driven decision modeling confronts two significant challenges: scarcity or constraints in data availability and inadequacy or bias in data quality~\cite{ref_40_9512658}. Limited or biased historical data originating from other agents frequently fails to accurately capture their genuine models. Moreover, inadequate or restricted data sampling hinders the development of comprehensive intent models, ultimately leading to a loss of information from the original historical decision sequences. Consequently, these factors can contribute to sub-optimal or non-generalizable decision models for the subject agent.

In this paper, we propose a novel data-driven framework to learn the true policy model of other agents from limited historical interaction data. We extract historical behavior characteristics to learn a new set of incomplete behavior models from interaction sequences. To enhance modeling accuracy, we adapt the Variational Autoencoder~(VAE) \cite{ref_41_9675815,ref_16_vae2021} to generate a collection of models that can encompass the true models of other agents from an incomplete set of policy trees. This new approach leverages information from incomplete policy trees that are often discarded or approximated by traditional methods. We develop a perplexity-based metric to quantify the likelihood of including true behavior models and select optimized top-$K$ behaviors from the large candidate set. Our contributions are summarized as follows:

\begin{itemize}
    \item We propose the VAE-based algorithm to address challenges in data-driven I-DIDs, resulting in VAE-enabled behaviors. This paves the way for future research on neural networks in MSDM problems.
    \item We develop the Zig-Zag One-Hot~(ZZOH) encoding and decoding technique, tailored specifically for policy trees, which empowers the VAE to efficiently handle both complete and incomplete policy trees as input data, thus enhancing its versatility.
    \item We analyze the quality of I-DID solutions using a novel perplexity-based metric, providing insights for algorithm fine-tuning and confidence in the performance.
    \item We conduct empirical comparisons to state-of-the-art I-DID solutions in two problem domains and investigate the potential of the novel algorithms.
\end{itemize}

The remainder of this paper is organized as follows. Section \ref{sec:related} reviews related works on solving I-DIDs. Section \ref{sec:background} provides background knowledge on I-DIDs and VAE. Section \ref{methods} presents our approach to generating behaviors of other agents in I-DIDs. Section \ref{experiments} shows the experimental results by comparing various I-DID solutions. Finally, we conclude the research and discuss future work in Section \ref{sec:final}.
\section{Related Works}
\label{sec:related}
This section reviews the current research and emerging trends in three key areas: neural computing-based decision-making modeling, modeling of unknown agent behaviors, and I-DID models, along with related data-driven decision-making approaches.

Neural computing techniques, particularly deep reinforcement learning~(DRL), have attained significant progress in addressing multi-agent decision-making challenges~\cite{ref_39_9119863,ref_12_Hazra2023,ref_13_Belle2024}. DRL, which combines deep learning and reinforcement learning, facilitates end-to-end learning control but encounters limitations in sparse rewards, limited samples, and multi-agent environments. Recent advancements in hierarchical, multi-agent, and imitation learning, along with maximum entropy-based methods, offer promising research directions~\cite{ref_15_9904958}. Investigations such as Kononov and Maslennikov's recurrent neural networks trained through reinforcement learning~\cite{ref_14_Kononov2023} and Zhang \emph{et al.}'s method for generating natural language explanations for intelligent agents' behavior~\cite{ref_10_zhang2023explaining}, demonstrate the potential of DRL. Additionally, MO-MIX and other DRL approaches enable multi-agent cooperation across diverse domains~\cite{ref_17_MO-MIX,ref_18_YANG2024109754,ref_19_le2022deep,ref_20_zhu2021deep,ref_37_9455523}.

To emulate human-level intelligence and handle unexpected agent behaviors, neuro-symbolic multi-agent systems are emerging. These systems leverage neural networks' ability to extract symbolic features from raw data, combined with sophisticated symbolic reasoning mechanisms. Techniques like agent-based models~(ABMs)~\cite{ref_11_9527397,ref_11_BERGER2024106003}, data-driven decision-making~\cite{ref_3_Buijsse2023,ref_2_Wang2023}, and large language models~(LLMs)~\cite{ref_9_10520238} are being explored. An \emph{et al.}~\cite{ref_4_AN2023} proposed a reinforcement learning and CNN-based approach for agents to self-learn behavior rules from data, promoting the integration of ABMs with data science and AI. Wang \emph{et al.}~\cite{ref_38_10172334} presented optimal decision-making and path planning for multi-agent systems in complex environments using mean-field modeling and reinforcement learning for unmanned aerial vehicles. Wason \emph{et al.}~\cite{ref_8_J10498880} and Nascimento \emph{et al.}~\cite{ref_7_Nascimento} explored integrating large language models into multi-agent systems, highlighting their human-like capabilities but also their distinctiveness, agent autonomy, and decision-making in dynamic environments.

Similarly, several I-DID solutions have relied solely on behavioral and value equivalences to constrain the model space for other agents, presuming that the true behaviors of those agents fall within the subject agent's modeling capabilities~\cite{ref_29_Zeng12:Exploiting, ref_30_DoshiGD20}. However, this approach falls short in fully accounting for unexpected or novel behaviors. Pan \emph{et al.}~\cite{ref_5_Pan2022} proposed a genetic algorithm-based framework that incorporates randomness into opponent modeling, thus generating novel behaviors for agents, thereby demonstrating the significant potential of evolutionary approaches.

In the context of data-driven I-DIDs~\cite{ref_6_Pan2021TowardDS}, the objective is to optimize multi-agent decision-making in uncertain environments, a significant challenge in AI research. Our work in this article pioneers a novel approach to modeling other agents in I-DIDs, exploring neural computing-based data-driven methods to approximate real agent behavior models from historical data in agent planning research. 

\section{Preliminary Knowledge}
\label{sec:background}
As we will adapt the VAE-based data generation methods to augment the I-DID model with novel metrics for other agents and generate novel behaviors, we provide background knowledge on I-DID and VAE.

\subsection{Background knowledge on I-DIDs}
The traditional influence diagram, designed for single-agent decisions, has evolved into the I-DID framework. This framework is a probabilistic graph tailored for interactive multi-agent decision-making under partial observability. From the agent $i$'s perspective, the I-DID predicts agent $j$'s actions, aiding agent $i$ in optimizing its decisions. We focus on agent $i$ and integrate agent $j$'s potential behaviors into the decision framework.

\begin{figure}[!t]
    \centering
    \includegraphics[width=0.48\textwidth]{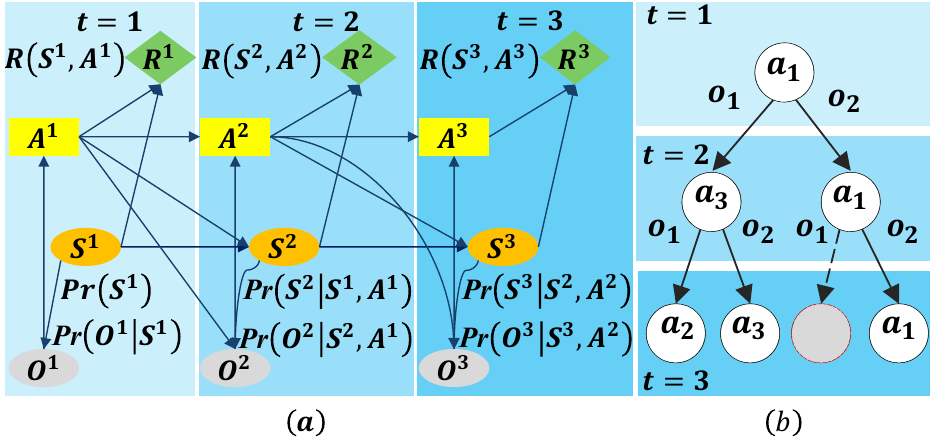}
    \caption{A dynamic influence diagram and its solutions:~($a$)the left is the dynamic influence diagram with three time steps and ($b$) the right is its solution represented as a policy tree. The  blocks with the same color in~($a$) and~($b$) belong to the same time slice.}
    \label{fig:DID}
\end{figure}

We introduce the dynamic influence diagram (DID) for a single agent in Fig.~\ref{fig:DID}(a). This DID represents the agent's decision process over three time steps, with solutions shown in Fig.~\ref{fig:DID}(b).
In the DID, ovals represent either chance nodes for environmental states ($S$) or observations ($O$); rectangles are decision nodes for the agent's actions ($A$); diamonds denote utility nodes capturing rewards ($R$).
At time $t$, the agent's decision ($A^t$) is influenced by the current observation ($O^t$) and the previous decision ($A^{t-1}$). The observation ($O^t$) depends on the previous decision ($A^{t-1}$) and the current state ($S^t$). The states ($S^t$) are partially observable and influenced by the previous states ($S^{t-1}$) and decision ($A^{t-1}$). Rewards ($R^{t-1}$) are determined by a utility function considering both the states and decisions.
Arcs model conditional probabilities among the connected nodes. For example, if $O^t$ is influenced by $A^{t-1}$ and $S^t$, the arrows from $A^{t-1}$ and $S^t$ to $O^t$ indicate this dependency.
Similarly, if $S^t$ is affected by $A^{t-1}$ and transitions from $S^{t-1}$, the arrows from $A^{t-1}$ and $S^{t-1}$ to $S^t$ show the state transition.
The DID demonstrates how the agent optimizes its decisions in the response to the changing environment. After defining the transition, observation, and utility functions, we use traditional inference algorithms to solve the model and get the agent's optimal policy. This policy is represented by a policy tree (see Fig.~\ref{fig:DID}(b)). At time $t=1$, the agent takes action $a_1$ and the subsequent actions based on observations. The paths in the tree correspond to the observations, while nodes represent the agent's chosen actions. In different domains, the number of the paths and node types may vary.
The policy tree, as a DID solution, encapsulates the model's optimal policy~(behavioral model). It's a full $k$-ary tree, where $k$ is the number of possible observations.
\begin{figure}[!t]
    \centering
    \includegraphics[width=0.4\textwidth]{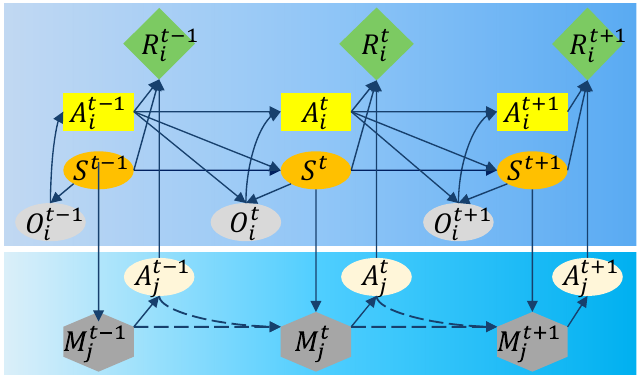}
    \caption{By extending DID model~(the blue part), the agent $i$ optimises its decision in the I-DID models with the blue part which models agent $j$'s decision making process.}
    \label{fig:I-DID}
\end{figure}

In the I-DID model, a hexagonal node, known as the model node $M_j$, dynamically extends the behavioral models of other agents in the influence diagram. This node contains potential behavioral models of agent $j$, which are then provided to the subject agent $i$, transforming the complex I-DID problem into a conventional DID. However, as the number of time slices increases, managing an extensive set of candidate models becomes intractable. Compression and pruning of model nodes are often necessary before integrating them into the DID, but this can result in nodes with similar behavioral models, leading to a limited set of optimal decisions for agent $i$.

In this article, to address the challenges posed by a large number of time slices and selecting vast candidate models, we use a variational autoencoder~(VAE) to learn from agent $j$'s historical trajectory. The new method generates a set of highly reliable policy trees, capturing key features of the trajectory. By embedding the potentially true behavioral model of agent $j$ into the I-DID, we enable agent $i$'s optimal policy to consider the most probable behaviors of agent $j$.

\subsection{Background knowledge on VAE}
Variational Autoencoders (VAEs)~\cite{ref_28_Kingma2013AutoEncodingVB,ref_35_Ding2021TheRF} are a type of unsupervised learning model that combines the capabilities of autoencoders with probabilistic generative models. Unlike traditional autoencoders~\cite{ref_27_Hinton06:Reducing}, which encode inputs into a fixed latent representation, VAEs map the inputs to a probability distribution in the latent space. 
The key idea behind VAEs is to encode input data into the parameters of a latent distribution, typically a Gaussian distribution. The encoder network predicts the mean ($\boldsymbol{\mu}_{\phi}$) and variance ($\boldsymbol{\sigma}_{\phi}^2$) of this distribution, while the decoder network reconstructs the original input from latent samples.
To enable differentiable sampling from the latent distribution, VAEs employ the reparameterization trick. This involves sampling random noise from a standard normal distribution and scaling it according to the predicted mean and variance to obtain a latent sample ($\textbf{z}\in \mathbb{R}^l$). 

The VAE's objective function consists of two parts: a reconstruction loss that measures the similarity between the original input and the reconstructed output, and a Kullback–Leibler~(KL) divergence that regularizes the latent distribution to be close to a prior distribution~(e.g., a standard normal distribution). Optimizing this objective function leads to encoder and decoder networks that can encode meaningful representations of the input data into the latent space.
Formally, for a given input $\textbf{x}$ from dataset $\mathcal{D}$, the VAE loss function is defined as:
\begin{equation}
\mathcal{L}_{\phi,\theta}^{vae}(\textbf{x}) = \mathcal{L}^{\text{KL}}_{\phi,\theta}(\mathbb{P}_{\phi}(\textbf{z}\mid \textbf{x}),\mathbb{P}_{\theta}(\textbf{z})) + \mathcal{L}_{\phi,\theta}^{\text{recon}}(\textbf{x})
\end{equation}
where $\mathcal{L}^{\text{KL}}_{\phi,\theta}$ is the KL divergence between the posterior distribution $\mathbb{P}_{\phi}(\textbf{z}\mid \textbf{x})$ and the prior distribution $\mathbb{P}_{\theta}(\textbf{z})$, and $\mathcal{L}_{\phi,\theta}^{\text{recon}}$ is the reconstruction loss.
Over the entire dataset $\mathcal{D}$, the VAE loss function is given by:
\begin{equation}
\mathcal{L}_{\phi,\theta}^{vae}(\mathcal{D}) = \sum_{\textbf{x} \in \mathcal{D}} \mathcal{L}_{\phi,\theta}^{vae}(\textbf{x})
\end{equation}
By minimizing the loss function through the techniques like a stochastic gradient descent, the VAE learns to encode inputs into a latent space that captures their essential features while enabling the generation of new data samples that follow the distribution of the original input data, where we can apply stochastic gradient descent~(SGD)~\cite{ref_35_Ding2021TheRF} to find the optimized parameters $\phi^*$ and $\theta^*$.

\section{Policy Tree Generation Based on VAE}
\label{methods}
Policy tree generation, often referred to as learning agents' behavioral models, involves the automated construction of a behavioral representation in the form of a policy tree from extensive multi-agent interaction data. The behavioral model aims to predict the actions an agent will take given observations from the environment, which is typically represented as a complete multi-fork tree, as depicted in Fig.~\ref{fig:DID}.

From the perspective of agent $i$, it is not necessary to understand how agent $j$ optimizes its behavior but rather to predict how agent $j$ will act. In the I-DID model, agent $i$ is provided with $m$ policy trees representing agent $j$'s behavior, where these trees shall be complete. However, two challenges arise:

\begin{itemize}
    \item Many policy trees generated directly from historical behavior sequences are not fully developed - they are incomplete.
    \item The generated policy trees may not accurately reflect the true behavioral model of agent $j$.
\end{itemize}

Addressing these issues is crucial for the effective utilization of the policy trees in predicting and understanding multi-agent interactions. To address the first issue, there are two traditional methods, but both have their limitations~\cite{rossijcai15}. Discarding policy trees with missing nodes is problematic because it risks excluding significant policy sequences that may contain valuable information from the original historical data. This approach is only feasible when data is abundant, but in most cases, data is limited.
The alternative method, based on statistical analysis, estimates the probability of each action in the current time slice using the actions from the previous time slice and the current observations. Then, a roulette wheel selection process is employed to randomly fill in the missing actions in the policy tree. However, this approach has its shortcomings as well. The actions at different levels in the policy tree represent distinct contexts. For instance, the action $a^4$ at time slice $t=4$ is influenced not only by the action $a^3$ and its corresponding observation, but also by the actions and observations at earlier time slices such as $t=1$ and $t=2$. Consequently, the high confidence in taking action $a^2$ at $t=2$ does not necessarily translate to the high probability of taking the same action at $t=4$.

To tackle the second issue of potential incompleteness in modeling agent $j$'s behaviors, we endeavor to gather a broader and more diverse collection of behavioral models. The greater the diversity within this set, the higher the probability that it will capture the full spectrum of agent $j$'s actual behavior patterns. This approach ensures that our models are comprehensive and representative, increasing their reliability and adaptability.

To that end, we develop a new framework capable of generating a comprehensive ensemble of policy trees from agents' interaction data. This framework not only reconstructs policy trees accurately but also guarantees the diversity within the resulting set. To tackle these challenges, we develop a policy generation method leveraging variational autoencoder techniques.

\subsection{Reconstructing an Incomplete Policy Tree}
\label{subsection:RIPT}
Given the limitation that agents often cannot store a large number of policy trees to navigate complex multi-agent interactive environments, the agent must repeatedly traverse the policy from the root down during interactions. In a simple term, from the subject agent's perspective, we can construct potentially incomplete policy trees for other agents solely based on a single sequence of interaction data.

To obtain a long sequence of agent $j$'s action-observations, we control the interaction between agent $i$ and the environment. This sequence is denoted as $h^L = (a^1o^1, a^2o^2, \ldots, a^t o^t, \ldots, a^L o^L)$, where $a^t \in A$ and $o^t \in \Omega$ represent an alternating series of actions and observations at time slice $t$. The parameter $L$ is the total number of time slices in this interaction.

\begin{figure}[htbp]
    \centering
    \includegraphics[width=0.5\textwidth]{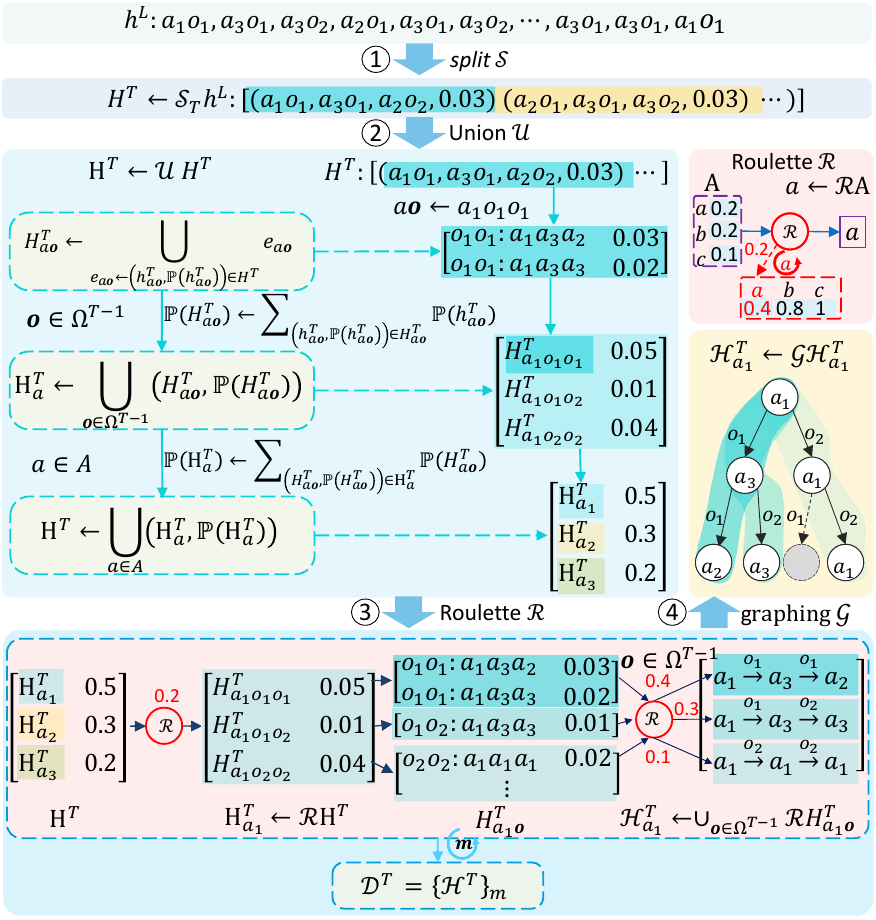}
    \caption{Reconstructing $m$ incomplete policy trees via four operators~($split$,$union$,$roulette$ and $graphying$) from behavior sequences.}
    \label{fig:ICOMTREE}
\end{figure}

Given the interaction data $h^L$, the depth of the policy tree~(denoted as $T$), and the number of agent $j$'s models~(denoted as $m$), we can reconstruct a set of possible incomplete policy trees $\bigcup \mathcal{H}^T$ of the fixed depth $T$~($T \ll L$) using four operators: $split$, $union$, $roulette$, and $graphing$. As illustrated in Fig.~\ref{fig:ICOMTREE}, these operators enable us to generate the desired set of the policy trees, and the implementation of these operators is described in Alg.~\ref{alg:OOP} of the Appendix.

   Utilizing the $split$ operator~(denoted as $\mathcal{S}$ in Fig.~\ref{fig:ICOMTREE} \textcircled{\small{1}}), we begin at the starting point and extract sequences of actions with a fixed length $T$, thereby forming policy paths that commence with a specific action $a$. We refer to such a path as $h^T_a$. If this policy path is aligned with a sequence of observations $\textbf{o}$, it can be denoted as $h^T_{a\textbf{o}}$. Concurrently, we compute the probability of each policy path, denoted as $\mathbb{P}(h^T_a) \leftarrow \#(h^T_a) \cdot T / L$. Here, $\#(h^T_a)$ representing the number of times $h^T_a$ appears in the entire sequence $h^L$. We repeat this operation until the end of the sequence, ultimately forming a set of policy paths $H^T$ denoted as $H^T \leftarrow \mathcal{S}_T(h^L)$.
    
    To obtain a set of policy trees from the set of policy paths $H^T$, we use the $union$ operator~(denoted as $\mathcal{U}$ in Fig.~\ref{fig:ICOMTREE} \textcircled{\small{2}}). First, we define a subset of policy paths that begin with the same action $a$ and observations $\textbf{o}$ as $H_{a\textbf{o}}^T = \bigcup_{(h^T_{a\textbf{o}},\mathbb{P}(h^T_{a\textbf{o}})) \in H^T} (h^T_{a\textbf{o}}, \mathbb{P}(h^T_{a\textbf{o}}))$. This subset contains all the policy paths that start with the same action $a$ and follow a specific sequence of observations $\textbf{o}$. We then compute the probability of this subset by summing up the probabilities of all the policy paths within it: $\mathbb{P}(H_{a\textbf{o}}^T) = \sum_{(h^T_{a\textbf{o}},\mathbb{P}(h^T_{a\textbf{o}})) \in H^T} \mathbb{P}(h^T_{a\textbf{o}})$. After collecting all the subsets $H_{a\textbf{o}}^T$ for every possible sequence of the observations $\textbf{o} \in \Omega^{T-1}$ following action $a$, we construct a larger set $\rm{H}_a^T$ that contains all the policy paths beginning with action $a$. This set is defined as $\rm{H}_a^T = \bigcup_{\textbf{o} \in \Omega^{T-1}} (H_{a\textbf{o}}^T, \mathbb{P}(H_{a\textbf{o}}^T))$ and its overall probability is computed as $\mathbb{P}(\rm{H}_a^T) = \sum_{\textbf{o} \in \Omega^{T-1}} \mathbb{P}(H_{a\textbf{o}}^T)$. Finally, for every action $a \in A$, we combine the sets $\rm{H}_a^T$ to form the complete set of policy paths $\rm{H}^T$, which is denoted as $\rm{H}^T = \bigcup_{a \in A} (\rm{H}_a^T, \mathbb{P}(\rm{H}_a^T))$. This set represents all possible policy trees of depth $T$ that can be constructed from the original interaction sequence $h^L$.
    
    To generate a set of incomplete policy trees $\mathcal{D}^T$ containing $m$ samples, we first apply the $roulette$ operator~(denoted as $\mathcal{R}$, Fig.~\ref{fig:ICOMTREE} \textcircled{\small{3}}) on the set of policy tree sets ${\rm{H}}^T$. This random selection process yields a single policy tree set $\rm{H}_a^T$ for a particular action $a$, represented as $\rm{H}_a^T \leftarrow \mathcal{R} {\rm{H}}^T$. Next, for each possible sequence of observations $\textbf{o} \in \Omega^{T-1}$ following action $a$, we apply the $roulette$ operator again to select a subset of policy paths from the corresponding subset $\emph{H}_{a\textbf{o}}^T$. This results in a collection of subsets, which is a possible policy tree~(denoted as $\mathcal{H}_a^T$), defined as $\mathcal{H}_a^T \leftarrow \bigcup_{\textbf{o} \in \Omega^{T-1}} \mathcal{R} \emph{H}_{a\textbf{o}}^T$. By repeatedly applying the $roulette$ operator in this manner, we generate $m$ incomplete policy tree collections, $\{\mathcal{H}^T\}_m$. The set is denoted as $\mathcal{D}^T$ and represents a diverse set of partially constructed policy trees, each of which starts with a randomly selected action $a$ and contains randomly sampled policy paths for each possible sequence of observations.
    
    When applying the $graphing$ operator~(denoted as $\mathcal{G}$, Fig.~\ref{fig:ICOMTREE} \textcircled{\small{4}}) to the set of policy paths $\mathcal{H}^T_a$ for the purpose of visualizing the corresponding policy tree, we represent this transformation as $\mathcal{H}^T_a \leftarrow \mathcal{G}\mathcal{H}^T_a$. Note that, although we use the same notation $\mathcal{H}^T_a$ to refer to both the policy tree and the set of policy paths rooted at $a$ with a length of time-slice $T$, the visualization process actually creates a representation of the policy tree in a graphical form. To clarify this distinction, we can explicitly state that $\mathcal{H}^T_a$ represents both the abstract policy tree structure and the concrete set of policy paths, while $\mathcal{G}\mathcal{H}^T_a$ represents the same policy tree but in a graphical representation generated by the $graphing$ operator $\mathcal{G}$.
    

As depicted in Fig.~\ref{fig:ICOMTREE}, the incomplete policy tree $\mathcal{H}^T_{a_1}$ is reconstructed from the historical behavior sequence $h^L$ through the application of the four operators: $split$, $union$, $roulette$, and $graphing$~(Fig.~\ref{fig:ICOMTREE} \textcircled{\small{1}}-\textcircled{\small{4}}). Additionally, Fig.~\ref{fig:DID} illustrates a scenario where, after agent $j$ observes $o_2$ in the first time slice and takes action $a_1$, it is unable to select a subsequent action if it encounters the observation $o_1$. Notably, some incomplete policy trees are inevitably created during the generation of agent $j$'s policy trees from historical data. These incomplete policy trees, though a special type of behavior model, differ from complete policy trees in that they lack specified actions for certain observations in certain time slices. However, the I-DID model requires a complete policy tree model for its solution methodology. Therefore, an incomplete policy tree cannot be directly utilized in I-DID. Thus, we propose a VAE-based method to convert incomplete policy trees into their complete counterparts, thereby enabling their utilization within the I-DID.
\subsection{Standard Policy Tree Generalization}

As mentioned earlier, discarding or randomly processing incomplete policy trees can result in a significant loss of crucial information from the original historical decision sequences. To address this, we develop a VAE-based approach for generating policy trees. This method effectively leverages both complete and incomplete policy trees, enabling the production of numerous new policy trees from just a few examples. By doing so, it maximizes the coverage of agent $j$'s true behaviors.

\begin{figure}[htbp]
    \centering
    \includegraphics[width=0.6\textwidth]{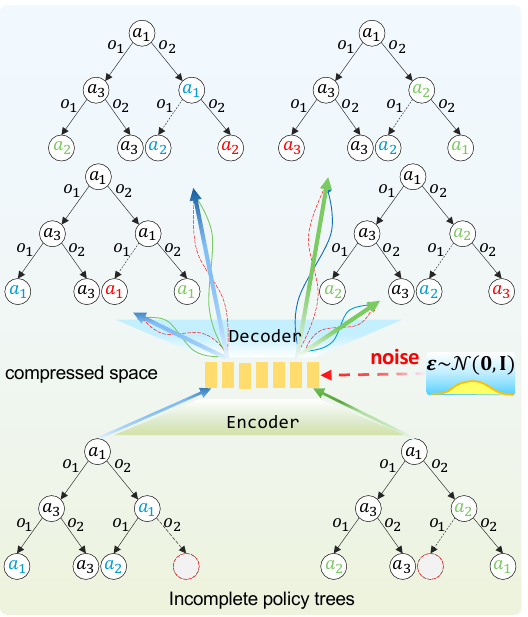}
    \caption{The principle of a VAE-based approach that leverages incomplete policy trees to generate diverse new trees, maximizing coverage of agent $j$'s true behaviors and enhancing the diversity of the overall policy tree collection.}
    \label{fig:vae_generation}
\end{figure}

In this manner, our approach solves two key problems faced by traditional methods in modeling agent $j$'s behaviors. As illustrated in Fig.~\ref{fig:vae_generation}, after training the VAE with a selection of incomplete policy trees, the model is capable of generating policy trees with varying degrees of deviation, adhering closely to the distribution of the historical data. This approach significantly enhances the diversity of the overall policy tree collection, leading to a more comprehensive representation of agent $j$'s behaviors.

Here, we introduce the Zig-Zag One-Hot~(ZZOH) encoding and decoding technique specifically designed for policy trees. This technique enables the VAE to effectively handle both the complete and incomplete policy trees as input and output data. Since ZZOH encoding prioritizes serializing the policy tree structure, nodes closer to the root have a greater impact on the overall decision-making effectiveness of the entire policy tree.

Therefore, we adapt the loss function of the VAE network to prioritize learning behaviors from nodes associated with earlier time slices. This approach helps capture the structural information embedded in the original policy tree. After reconstructing and filtering a large number of policy trees using the VAE, we utilize the Measurement of diversity with frames~(MDF) and information confusion degree~(IDF) to select the top-$K$ policy trees from the generated set.

\begin{figure*}[htbp]
    \centering
    \includegraphics[width=0.8\textwidth]{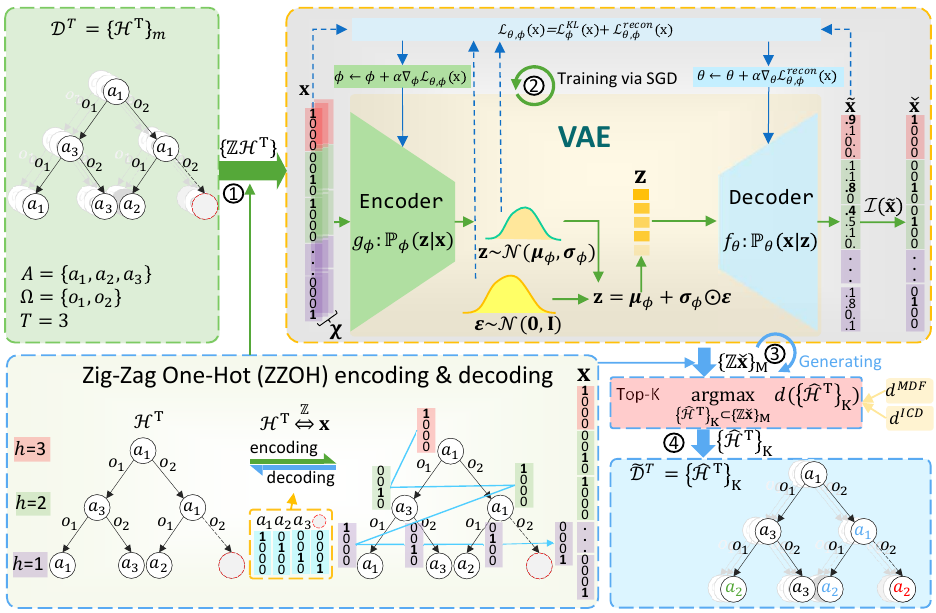}
    \caption{The VAE network creates new policy trees. With the Zig-Zag One-Hot encoding-decoding method designed for policy trees, VAE can handle both complete and incomplete trees, emphasizing learning from earlier nodes. By using MDF and IDF, we pick the most diverse and reliable top-$K$ trees, matching historical agent patterns closely.}
    \label{fig:vae_network}
\end{figure*}

Empirically, we verify that the behavior model of agent $j$ generated using the VAE closely aligns with the distribution of agent $j$'s historical data. Furthermore, we compare the diversity and credibility of the top-$K$ policy trees. We have described the detailed process of generating more diverse policy trees using the VAE model in Fig.~\ref{fig:vae_network}.

\subsubsection{Zig-Zag One-Hot encoding \&\ decoding}
\label{subsection:ZZOH}
To serialize the policy tree and make it compatible with the VAE network, we introduce the Zig-Zag One-Hot~(ZZOH) encoding and decoding technique. This approach effectively transforms both complete and incomplete policy trees into column vectors that can be utilized by the VAE network.

Considering some empty nodes in incomplete policy tree, we enlarge the action space $A_j=\{a_1,a_2,...a_{|A_j|}\}$ of agent $j$ with empty action $a_0$ , denoted as $\Tilde{A}_j=\{a_0,a_1,a_2,...a_{|A_j|}\}$. Then, we use one-hot encoding to encode each action of the enlarged action space $\Tilde{A}_j$ into a binary code with the length $|\Tilde{A}_j|$ and generate an encoding set of action space, denoted as $\Tilde{A}_j^c$. For example, given the action space $A_j=\{a_1,a_2,a_3\}$, we have enlarged the action space $\Tilde{A}_j=\{a_0,a_1,a_2,a_3\}$, and the one-hot encoding set of enlarged action space \[  
\Tilde{A}_j^c = \begin{matrix}  
a_0 & a_1 & a_2 & a_3 \\  
1 & 0 & 0 & 0 \\  
0 & 1 & 0 & 0 \\  
0 & 0 & 1 & 0 \\  
0 & 0 & 0 & 1  
\end{matrix}  
\], and we have $\Tilde{A}_j^c[a_0]=[1;0;0;0]$.
By encoding the action space, we define an operator $\mathcal{Z}$ for encoding/decoding policy trees to/from binary codes. This lossless transformation into a vector representation preserves the tree's structural information, enabling efficient processing and analysis.

Thus, for given a policy tree $\mathcal{H}^T$, we encode each action node into a binary code using the encoding set of enlarged action space, and then concatenate the binary codes into a column vector $x$ in a zigzag order, denoted as $\mathcal{Z}\mathcal{H}^T$. Wherein the numerical information indicates the node action information, and the node position information indicates the sequence information in front of the node and the current observation result.
We repeat this operation to encode the policy tree in the given policy tree set $\mathcal{D}^T$ and form the dataset $X$ for training and testing the VAE network, which is denoted as $\mathcal{Z}\mathcal{D}^T$, as shown in Fig.~\ref{fig:vae_network} \textcircled{\small{1}}. We also present the implementation of the operators in Alg.~\ref{alg:ZZOH}  of the Appendix.

\subsubsection{Tree Loss Function in VAE}
The intricate structure of policy trees encapsulates critical information, including action values, sequential order, and their relative importance. Action nodes closer to the root exercise greater influence on the overall decision-making effectiveness of the tree, shaping the trajectory to all the leaf nodes. However, this structural information is not fully captured in serialized tree representations.

To leverage serialized policy tree data for VAE training and better capture structural nuances, we propose a modified VAE loss function. This refinement prioritizes the behavior learning from nodes with smaller time slices, emphasizing the structural information embedded in the original policy tree. Effectively, this approach assigns higher weights to action nodes closer to the root, ensuring that their significance is reflected in the VAE learning.

Given the prior distribution over latent variables as a centered isotropic multivariate Gaussian:
${\mathbb{P}_{\theta}(\textbf{z})} = \mathcal{N}(\textbf{z};\textbf{0},\textbf{I})$, we define the likelihood function ${\mathbb{P}_{\theta}(\textbf{x}\mid \textbf{z})}$ as a multivariate Bernoulli distribution (for binary data) where the distribution parameters are determined from $\textbf{z}$ using a fully-connected neural network with a single hidden layer. 
Since the true posterior ${\mathbb{P}_{\theta}(\textbf{z}\mid \textbf{x})}$ is often intractable, we employ a variational approximate posterior ${\mathbb{P}_{\phi}(\textbf{z}\mid \textbf{x})}$ to approximate it. Specifically, we let the variational approximate posterior be a multivariate Gaussian with a diagonal covariance structure:
${\mathbb{P}_{\phi}(\textbf{z}\mid \textbf{x})} = \mathcal{N}(\textbf{z};\boldsymbol{\mu}_{\phi}, \boldsymbol{\sigma}_{\phi}^2\textbf{I})$.
where $\boldsymbol{\mu}_{\phi}$ and $\boldsymbol{\sigma}_{\phi}^2$ are the mean and variance vectors predicted by the encoder network parameterized by $\phi$.
To enable differentiable sampling from this variational posterior, we utilize the reparameterization trick. Specifically, we sample a random noise vector $\boldsymbol{\varepsilon}$ from a standard normal distribution $\mathcal{N}(\textbf{0},\textbf{I})$ and then transform it according to the predicted mean and variance:
$\textbf{z} = \boldsymbol{\mu}_{\phi} + \boldsymbol{\sigma}_{\phi} \odot \boldsymbol{\varepsilon}$.
This re-parameterization allows gradients to flow through the sampling process, enabling the use of gradient-based optimization techniques to train the VAE.
 \begin{equation}
 \begin{split}
\boldsymbol{\mu}_{\phi},\boldsymbol{\sigma}_{\phi}&= Encoder_{\phi}(\textbf{x})\\
\boldsymbol{\varepsilon} \sim  {\mathbb{P}(\boldsymbol{\varepsilon}) } &= \mathcal{N}(\boldsymbol{\varepsilon};\textbf{0},\textbf{I})\\
\textbf{z}&= \boldsymbol{\mu}_{\phi}+ \boldsymbol{\sigma}_{\phi} \odot \boldsymbol{\varepsilon}
\end{split}
\end{equation}
We have a the decoder of VAE for Bernoulli data:
 \begin{equation}
 \begin{split}
\Tilde{\textbf{x}} &= Decoder_{\theta}(\textbf{z})\\
 \log{\mathbb{P}_{\theta}(\textbf{x}\mid\textbf{z} )} 
&= \sum_{k=1}^{|\textbf{x}|}\log{\mathbb{P}_{\theta}(\textbf{x}_k\mid\textbf{z} )}\\
&= \sum_{k=1}^{|\textbf{x}|}{Bernoulli(\textbf{x}_k;\Tilde{\textbf{x}}_k )}\\
&= \sum_{k=1}^{|\textbf{x}|}{\textbf{x}_k\log \Tilde{\textbf{x}}_k+(1-\textbf{x}_k)\log (1-\Tilde{\textbf{x}}_k)}\\
&= \ell(\textbf{x}, \Tilde{\textbf{x}})\\
\end{split} 
\end{equation}
where $\ell(\textbf{x}, \Tilde{\textbf{x}})$ denotes the Binary Cross-Entropy Loss (BCELoss)~\cite{ref_35_Ding2021TheRF} due to the VAE decoder for Bernoulli data, aka the basic construction error.

To utilize serialized data from the policy tree to train a VAE network and better represent the structural information and data distribution characteristics of the original policy tree, we propose the decoder of VAE for Bernoulli data with tree weights.
 \begin{equation}
 \begin{split}
\ell^{tree}(\textbf{x}, \Tilde{\textbf{x}}) &= \sum_{k=1}^{|\textbf{x}|}{\textbf{w}(k)(\textbf{x}_k\log \Tilde{\textbf{x}}_k+(1-\textbf{x}_k)\log (1-\Tilde{\textbf{x}}_k))}\\
\end{split} 
\end{equation}
where $\textbf{w}(k) = \log {(1+h(\lfloor \frac{k-1}{N} \rfloor+1))}$, $N = |\textbf{x}|/(|A_j|+1)$,$\forall \Tilde{\textbf{x}}_i \in \Tilde{\textbf{x}}: 0\leq \Tilde{\textbf{x}}_i \leq 1 $. $h(n)$ is the height of a node in the policy tree, which can be calculated as:
\begin{equation}
h(n) = T-c+1, if ~n\in [\frac{|\Omega|^{c-1}-1}{|\Omega|-1}+1,\frac{|\Omega|^{c}-1}{|\Omega|-1}],c\in \mathbb{R}^+
\end{equation}
Then we have the loss function of VAE: 
\begin{equation}
 \begin{split}
\mathcal{L}_{\theta,\phi}^{\text {recon }}(\textbf{x}) &=-\mathbb{E}_{\textbf{z}\sim \mathbb{P}_{\phi}(\textbf{z}\mid \textbf{x})}{\log(\mathbb{P}_{\theta}(\textbf{x}\mid\textbf{z}  ))}\\
&=-\mathbb{E}_{\boldsymbol{\varepsilon} \sim \mathcal{N}(\boldsymbol{\varepsilon};\textbf{0},\textbf{I})}{\log(\mathbb{P}_{\theta}(\textbf{x}\mid\boldsymbol{\mu}_{\phi}+ \boldsymbol{\sigma}_{\phi} \odot \boldsymbol{\varepsilon} ))}\\
&= -\frac{1}{2n_s} \sum_{l=1}^{n_s} \ell^{tree}(\textbf{x}, \Tilde{\textbf{x}}^{(l)}) 
\end{split}
\end{equation}
\begin{equation}
 \begin{split}
\mathcal{L}_{\theta,\phi}^{\text {KL}}(\textbf{x};\phi)&= {\mathcal{L}_{\text{KL}}(\mathbb{P}_{\phi}(\textbf{z}\mid \textbf{x}),\mathbb{P}_{\theta}(\textbf{z} ))}\\
&= {{D}_{\text{KL}}(\mathcal{N}(\textbf{z};\boldsymbol{\mu}_{\phi}, \boldsymbol{\sigma}_{\phi}^2\textbf{I})\mid \mid \mathcal{N}(\textbf{z};\textbf{0},\textbf{I}))}\\
&= \frac{1}{2} \sum_{k=1}^{|\boldsymbol{\mu}_{\phi}|}\left(1+\log \boldsymbol{\sigma}_{\phi,k}^2-\boldsymbol{\mu}_{\phi,k}^2-\boldsymbol{\sigma}_{\phi,k}^2\right)\\
&=\mathcal{L}_{\phi}^{\text {KL}}(\textbf{x};\phi)\\
\end{split}
\end{equation}
\begin{equation}
\mathcal{L}_{\theta,\phi}(\textbf{x}) = \mathcal{L}_{\theta,\phi}^{\text {recon }}{(\textbf{x})}+\mathcal{L}_{\phi}^\text{K L}{(\textbf{x})}\\
\end{equation}
\begin{algorithm}[htbp]
\caption{Learning a tree-loss based VAE network through SGD~($@\rm{VAE}$)}
\label{alg:alg1}
\KwData{Dataset $\mathcal{X} = \{\textbf{x}\}_{m}$, learning rate $\alpha$, batch size $n_b$, sampling size $n_s$.
}
\KwResult{Learned network $\text{Net}_{\theta,\phi}^{vae}(\cdot)$ with parameters $\theta$ and $\phi$}
\textbf{Initialize:} Variational network parameters $\theta$ and $\phi$ randomly \\
\Repeat
{Convergence}
{
    Sample a batch of $n_b$ data $\Tilde{\mathcal{X}}=\{x\}_{n_b}$ from $\mathcal{X}$\\  
    \For{$\textbf{x}\in \Tilde{\mathcal{X}}$}  {
        $\boldsymbol{\mu},\boldsymbol{\sigma}\leftarrow Encoder_{\phi}(\textbf{x})$\\
        \For{$l\in \{1,2,\dots,n_s\}$}  {
            Generate noise $\boldsymbol{\varepsilon} \sim \mathcal{N}(\textbf{0},\textbf{I})$\\
            $\textbf{z}\leftarrow \boldsymbol{\mu}+ \boldsymbol{\sigma} \odot \boldsymbol{\varepsilon}$\\
             $\Tilde{\textbf{x}}^{(l)} \leftarrow Decoder_{\theta}(\textbf{z})$\\
        }
    Compute the loss function:\\
    $\mathcal{L}_{\theta,\phi}^{\text {recon }}(\textbf{x})\leftarrow  -\frac{1}{2n_s} \sum_{l=1}^{n_s} \ell^{tree}(\textbf{x}, \Tilde{\textbf{x}}^{(l)})$\\
    $\mathcal{L}_{\phi}^{\text{KL}}(\textbf{x}) \leftarrow\frac{1}{2} \sum_{k=1}^{|\boldsymbol{\mu}_{\phi}|}\left(1+\log \boldsymbol{\sigma}_{\phi,k}^2-\boldsymbol{\mu}_{\phi,k}^2-\boldsymbol{\sigma}_{\phi,k}^2\right)$\\
    }

    Compute the gradients of the loss function with respect to the network parameters: $\nabla_\theta \sum_{\textbf{x}} \mathcal{L}_{\theta,\phi}^{\text {recon }}{(\textbf{x})}$\\
    $\nabla_\phi \sum_{\textbf{x}}\left(\mathcal{L}_{\theta,\phi}^{\text {recon }}{(\textbf{x})}+\mathcal{L}_{\phi}^{\text{KL}}{(\textbf{x})}\right)$\\ 
    Update parameters using gradient descent: \\
    $\theta \leftarrow \theta-\alpha \nabla_\theta \sum_{\textbf{x}} \mathcal{L}_{\theta,\phi}^{\text {recon }}{(\textbf{x})}$ \\
    $\phi \leftarrow \phi-\alpha \nabla_\phi \sum_{\textbf{x}}\left(\mathcal{L}_{\theta,\phi}^{\text {recon }}{(\textbf{x})}+\mathcal{L}_{\phi}^\text{K L}{(\textbf{x})}\right)$
}
\end{algorithm} 
where $n_s$ is the batch learning sampling size, $\Tilde{\textbf{x}}=f_\theta(g_\phi(\textbf{x}))$ is the output vector of VAE network, the operator $\odot$ is a element-wise product of two vector and removes all the zero-elements. 

Subsequently we apply stochastic gradient descent algorithm~(SGD) with the batch learning to train the VAE network, minimizing the modified reconstruction loss. First, we compute the gradients of the loss function with respect to the network parameters: $\nabla_\theta \sum_{\textbf{x}} \mathcal{L}_{\theta,\phi}^{\text {recon }}{(\textbf{x})}$ and
    $\nabla_\phi \sum_{\textbf{x}}\left(\mathcal{L}_{\theta,\phi}^{\text {recon }}{(\textbf{x})}+\mathcal{L}_{\phi}^{\text{KL}}{(\textbf{x})}\right)$. Then, we update parameters using gradient descent:$\theta \leftarrow \theta-\alpha \nabla_\theta \sum_{\textbf{x}} \mathcal{L}_{\theta,\phi}^{\text {recon }}{(\textbf{x})}$ and
    $\phi \leftarrow \phi-\alpha \nabla_\phi \sum_{\textbf{x}}\left(\mathcal{L}_{\theta,\phi}^{\text {recon }}{(\textbf{x})}+\mathcal{L}_{\phi}^\text{K L}{(\textbf{x})}\right)$. After the training process is completed, the original policy tree is reconstructed to expand and obtain the complete policy tree, as shown in Fig. ~\ref{fig:vae_network}. During the training phase of VAE, we will train the VAE network parameters  to learn the distribution of the behaviors from historical policy trees that may have some incomplete policy trees. In the testing, we reconstruct the input policy tree to obtain a complete policy tree, including the selection probabilities for each node.

We outline the VAE training in Alg.~\ref{alg:alg1}. We initialize the VAE parameters $\theta$ and $\phi$ and choose a data batch of size $n_b$ (lines 1-2). Then we encode each datum to get the compressed variable statistics and decode with the noise to reconstruct~(lines 5-10). Subsequently we adjust and compute the VAE loss~(lines 11-13), calculate the gradients, update the parameters, and repeat the procedures until the parameter values become stable~(lines 15-20).
After the network parameters are finalized, we input the given policy trees $\mathcal{H}^T$ into the trained VAE network. This generates a new vector $\check{\mathbf{x}}$ representing a policy tree as outputs. Applying the One-hot encoding operator $\mathcal{I}(\Tilde{\mathbf{x}})$ transforms a probability vector to one-hot binary vector for each action node, which means selecting the action with the highest probability in the corresponding chosen vector for the node, presented in Alg.~\ref{alg:one-hot}. Then, we apply the ZOOH decoding operator to produce a policy tree $\mathcal{Z}\check{\mathbf{x}}$, ensuring that the resulting policy tree is complete and valid. We repeat this process until we obtain $\mathrm{M}$ complete policy trees~(Fig.~\ref{fig:vae_network} \textcircled{\small{3}}), denoted as $\check{\mathcal{D}}^T = \{\mathcal{Z}\check{\mathbf{x}}\}_{\mathrm{M}}$.

\subsection{Evaluation Index}
After reconstructing a significant number of policy trees, we need to filter out those that do not accurately represent agent $j$'s true behavior. To achieve this, we leverage the measurement of diversity with frames~(MDF)~\cite{ref_5_Pan2022} and introduce a novel metric called information confusion degree~(ICD). These metrics are to select the top-$K$ policy trees from the generated set.

MDF evaluates the diversity across both the vertical and horizontal dimensions: the vertical refers the variations in behavior sequences within each policy tree, and the horizontal explores the sequence differences across all potential observations at a time step.
\begin{equation}
\operatorname{d}\left(\{\mathcal{\hat{H}}^T\}_K\right)^{\rm{MDF}}=\sum_{t=1}^T \frac{\operatorname{Diff}\left(h_t\right)+\operatorname{Diff}\left(H_t\right)}{\left|\Omega_j\right|^{t-1}}
\label{MDF}
\end{equation}
where $Diff(h_t)$ and $Diff(H_t)$ count the different sequences $h_t$ and sub-trees~(frames) $H_t$ respectively in $\mathcal{D}^T_{K} = \{\mathcal{\hat{H}}^T\}_K$. $|\Omega_j|$ denotes the number of agent $j$'s observations. 

Additionally, we introduce ICD to evaluate the reliability of the policy tree data. This metric guarantees a high confidence for each node in the reconstructed tree. In the VAE-generated tree, nodes act as classifiers, choosing the most probable action. A small margin between the top action's probability and the rest leads to high information confusion, indicating unclear or unreliable decisions. We want the ICD value of the policy tree to reach the maximum value, and select the top-$K$ tree as the output. 
\begin{equation}
\operatorname{d}\left(\{\mathcal{\hat{H}}^T\}_K\right)^{\rm{ICD}}= - \sum_{\mathcal{\hat{H}}^T\in \{\mathcal{\hat{H}}^T\}_K}^{}{\sum_{n=1}^{N}{log(1+h(n)) * p_{n} log(p_{n})}}
\label{ICD}
\end{equation}
where $p_{n}$ is the probability of action under corresponding observation and $N$ is the number of action nodes. 

By selecting the top-$K$ policy trees, we aim for the maximum diversity or minimal information confusion among the generated trees. Since MDF captures the diversity of the overall behaviors and the ICD of these trees reflects their true distribution in the historical data, we conduct the top-$K$ selection using both MDF and ICD.
\begin{equation}
\operatorname{\{\mathcal{\hat{H}}^T\}_K} =\argmax_{\{\mathcal{\hat{H}}^T\}_K\subset\check{\mathcal{D}}^T} {d}\left(\{\mathcal{\hat{H}}^T\}_K\right)
\label{topk}
\end{equation}
where ${d}\left(\cdot \right)$ is calculate by  either  Eq.~\ref{ICD} or Eq.~\ref{MDF}.

\subsubsection{The Framework of the VAE-based Policy Tree Generation}
\label{subsection:VEB}
We elaborate the generation of the diverse policy trees using the VAE in Alg.~\ref{alg:alg2}. As depicted in Fig.~\ref{fig:ICOMTREE} and Fig.~\ref{fig:vae_network}, the key steps are as follows.

Initially, the process begins by reconstructing a set of incomplete policy trees, denoted as $\mathcal{D}^T$, from an agent $j$'s interactive history $h^L$. This reconstruction is achieved through a series of operations including $split$, $union$, $roulette$, and $graphying$, which serve to extract and organize policy trees from the historical data~(lines 1-9, Fig.~\ref{fig:ICOMTREE} \textcircled{\small{1}}-\textcircled{\small{4}}).

Subsequently, a ZZOH encoding operator is applied to each policy tree in $\mathcal{D}^T$, converting them into a binary vector representation. This encoded dataset, denoted as $X$, is then used to train and test the VAE network~(line 10, Fig.~\ref{fig:vae_network} \textcircled{\small{1}}). Utilizing stochastic gradient descent~(SGD) and a tree-specific loss function, the VAE network learns to capture the latent representations of the policy trees~(line 11, Fig.~\ref{fig:vae_network} \textcircled{\small{2}}).

Once the VAE network has been trained, it is ready to generate more complete policy trees~(lines 12-18, Fig.~\ref{fig:vae_network} \textcircled{\small{3}}). This is achieved by randomly selecting a binary vector from the original dataset $\mathcal{X}$ and feeding it into the VAE network. The network then generates a new binary vector, which is decoded using One-hot encoding and the ZZOH decoding operator to reconstruct a complete policy tree~(lines 14-17). This process is repeated until a desired number of complete policy trees are obtained~(lines 13-18).

Finally, from the generated set of policy trees $\check{\mathcal{D}}^T$, the most diverse or reliable behaviors are selected using metrics such as the MDF and ICD. These metrics allow us to identify the $K$ policy trees that exhibit the highest degree of diversity or reliability, enabling the selection of behaviors that are the best suited for a given task or scenario~(line 19, Fig.~\ref{fig:vae_network} \textcircled{\small{4}}).

\begin{algorithm}[htbp]
\caption{VAE-Enabled behaviors~($@VEB$)}
\label{alg:alg2}
\KwData{agent $j$'s interactive history $h^L$, learning rate $\alpha$, action set $A$ and observation set $\Omega$, the number of policy trees $m$, the number of generated policy trees $\rm{M}$, the evaluation index function $d(\cdot)$, the parameter of top-$K$ selection function $K$}
\KwResult{New $K$ policy trees $\operatorname{\{\mathcal{\hat{H}}^T\}_K}$}
${H^T}  \leftarrow \mathcal{S}_{T}{ h^L}$.~$\triangleleft$\emph{Applying $split$ operator}\;
$\rm{H}^\emph{T}\leftarrow  \mathcal{U} \emph{H}^\emph{T}$.~$\triangleleft$\emph{Applying $union$ operator}\;
$\mathcal{D}^T \leftarrow \emptyset$\\
\For {$k \in  \{1,2,\cdots m\} $}{
    $\rm H^T_a\leftarrow \mathcal{R} \rm{H}^\emph{T}$.~$\triangleleft$\emph{Applying $roulette$ operator}\;
    $\mathcal{H}^T_a\leftarrow \cup_{\textbf{o}\in \Omega^{T-1}}\mathcal{R} {\emph{H}_{a\textbf{o}}^\emph{T}}$.~$\triangleleft$\emph{Applying $roulette$ operator}\;
    $\mathcal{H}^T_a \leftarrow \mathcal{G} \mathcal{H}^T_a$.~$\triangleleft$\emph{Applying $graphing$ operator}\;
    $\mathcal{D}^T \leftarrow \mathcal{D}^T \cup \mathcal{H}^T_a$
}
$\mathcal{X}\leftarrow \cup_{\mathcal{H}^T\in \mathcal{D}^T}{\mathcal{Z}\mathcal{H}^T}$.~$\triangleleft$\emph{Applying ZZOH operator}\;
$\text{Net}_{\theta,\phi}^{vae}(\cdot)\leftarrow  \text{VAE}(\mathcal{X})$.~$\triangleleft$\emph{Learning VAE with tree loss}\;
$\check{\mathcal{D}}^T \leftarrow \emptyset$\\
\For {$k \in  \{1,2,\cdots \rm{M}\} $}{
    $\textbf{x}\leftarrow \mathcal{R} \mathcal{X}$.~$\triangleleft$\emph{Applying $roulette$ operator}\;
    $\Tilde{\textbf{x}}\leftarrow \text{Net}_{\theta,\phi}^{vae}(\textbf{x})$.~$\triangleleft$\emph{Generate data via VAE}\;
     $\check{\textbf{x}}\leftarrow \mathcal{I}(\Tilde{\textbf{x}})$.~$\triangleleft$\emph{Applying One-hot encoding}\;
     $\check{\mathcal{D}}^T \leftarrow \check{\mathcal{D}}^T \cup \mathcal{Z}\check{\textbf{x}}$ ~$\triangleleft$\emph{Applying ZZOH operator}\;
}
$\operatorname{\{\mathcal{\hat{H}}^T\}_K} \leftarrow \argmax_{\{\mathcal{\hat{H}}^T\}_K\subset\check{\mathcal{D}}^T} {d}\left(\{\mathcal{\hat{H}}^T\}_K\right)$~$\triangleleft$\emph{Applying top-$K$ operator}\;
\end{algorithm}

\section{Experimental Results}
\label{experiments}

We improve the I-DID by incorporating the VAE-based method to generate and select the diverse and representative behaviors for agent $j$. We conduct the experiments in two well-studied multi-agent problem domains: the multi-agent tiger problem~(Tiger)  and the multi-agent unmanned aerial vehicle~(UAV) problem~\cite{ref_42_Zeng2012Improved,ref_43_Conroy2015,ref_44_Doshi2009,ref_26_Zeng2016Approximating}. All the experiments are performed on a Windows 10 system with an 11-th Gen Intel Core i7-6700 CPU @ 3.40GHz~(4 cores) and 24GB RAM.

In the multi-agent context, we focus on a  general scenario with two agents, considering agent $i$ as the subject and constructing an I-DID model for the problem domains. The $M_j$ nodes in this model represent the behavioral models of agent $j$, as shown in Fig.~\ref{fig:I-DID}. Our proposed approach optimizes the I-DID model by providing agent $i$ with a set of agent $j$'s representative behaviors. We compare seven algorithms for solving the I-DID model in the experiments:
\begin{itemize}
    \item The classic I-DID algorithm, which relies solely on known models $M$ to expand the model nodes, assuming the true behavior of agent $j$ lies within these nodes~\cite{ref_44_Doshi2009}.
    \item  The genetic algorithm based algorithm~(IDID-GA) generates agent $j$’s behaviors through genetic operations~\cite{ref_45_mc21,ref_5_Pan2022} .
    \item The IDID-MDF~\cite{ref_5_Pan2022} and IDID-VAE-MDF algorithms, which utilize VAE methods to generate new data from agent $j$'s historical data, and then employ the MDF metric to select the top-$k$ behaviors of agent $j$.
    \item The IDID-Random, IDID-VAE-MDF, and IDID-VAE-ICD algorithms, which differ in how they select the top-$K$ behaviors from the policy trees generated by VAE. IDID-Random selects the behaviors randomly, while IDID-VAE-MDF and IDID-VAE-ICD utilize MDF and ICD respectively and IDID-VAE-BCELoss replaces the proposed tree loss with the traditional BCELoss.
\end{itemize}

All seven algorithms~(IDID, IDID-MDF, IDID-VAE-MDF, IDID-GA,IDID-Random, IDID-VAE-ICD and IDID-VAE-BCELoss) use the same underlying algorithm to solve the I-DID model, differing only in how they expand the model nodes with candidate models for agent $j$.

To evaluate the algorithms, we consider the average rewards received by agent $i$ during interactions with agent $j$. We randomly select one behavior model of agent $j$ as its true model from the set of $M_j$ nodes. Agent $i$ then executes its optimal I-DID policy, while agent $j$ follows the selected behavior. In each interaction, agent $i$ accumulates rewards based on the action results at each time step over the entire planning horizon $T$. We repeat the interaction 50 times and compute the average reward for agent $i$.


\subsection{Multi-agent tiger problems}

\begin{figure}[!t]
    \centering
    \includegraphics[width=0.45\textwidth]{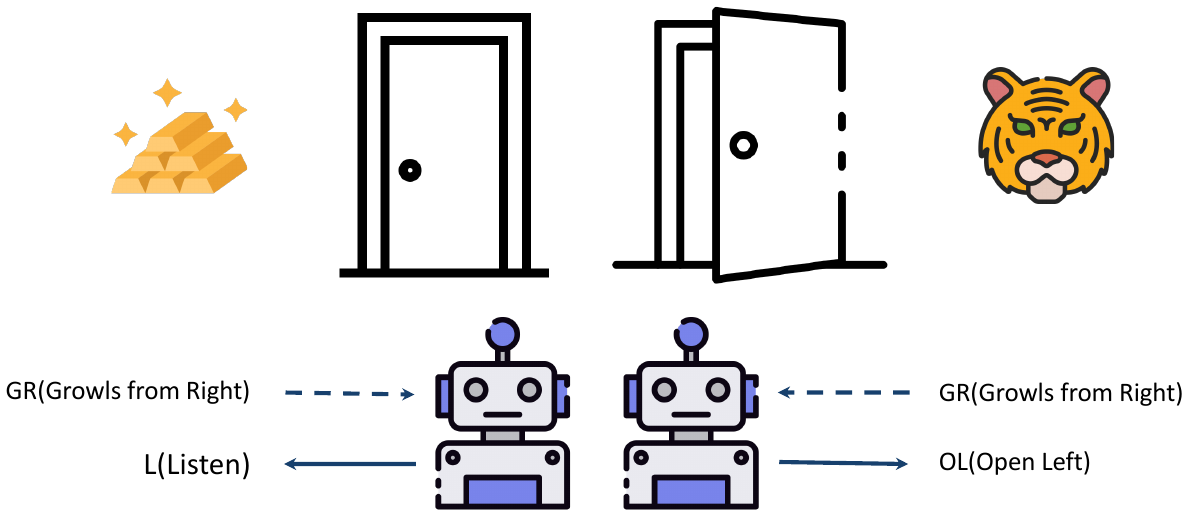}
    \caption{A multi-agent tiger problem, being simplified to the two-agent version, where modeling agent $j$'s behavior is to optimize agent $i$'s decision. The problem specification follows: $|S|=2$,$|A_i|=|A_j|=3$,$|\Omega_i|=6$, and $|\Omega _j|=2$.}
    \label{fig:tiger}
\end{figure}

The two-agent tiger problem serves as a canonical benchmark for assessing the performance of multi-agent planning models. 
In Fig.~\ref{fig:tiger}, agent $i$ and agent $j$ must make a choice between opening the right/left door~(OR/OL) or listening~(L) to determine the tiger's position, given limited visibility.
If an agent opens the door with gold behind it, it will take the gold; however, if the tiger is behind the door, the agent will be eaten. Meanwhile, if both agents simultaneously open the door with the gold, they will share the reward equally. The agents' decisions are influenced by their observations, which may not always be accurate. For instance, a squeak emanating from a door could be mistaken for another agent's voice, the tiger's growl, or a misinterpretation of the sound.

\subsubsection{Diversity and Measurements}
\begin{figure}[htbp]
\centering
\subfigure[T = 3]{
\includegraphics[width=0.47\linewidth]{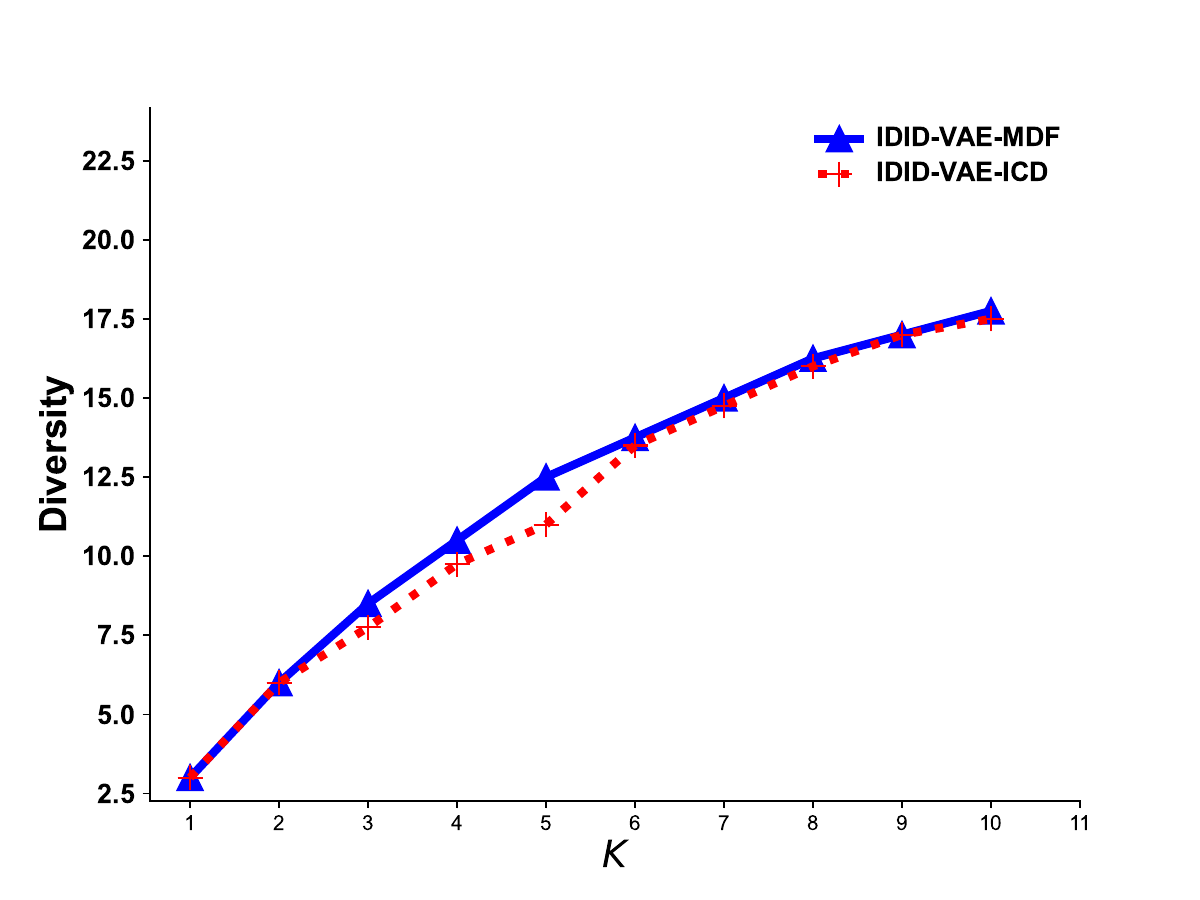}
\label{fig:tiger3_div}}
\hfil
\subfigure[T = 4]{
\includegraphics[width=0.47\linewidth]{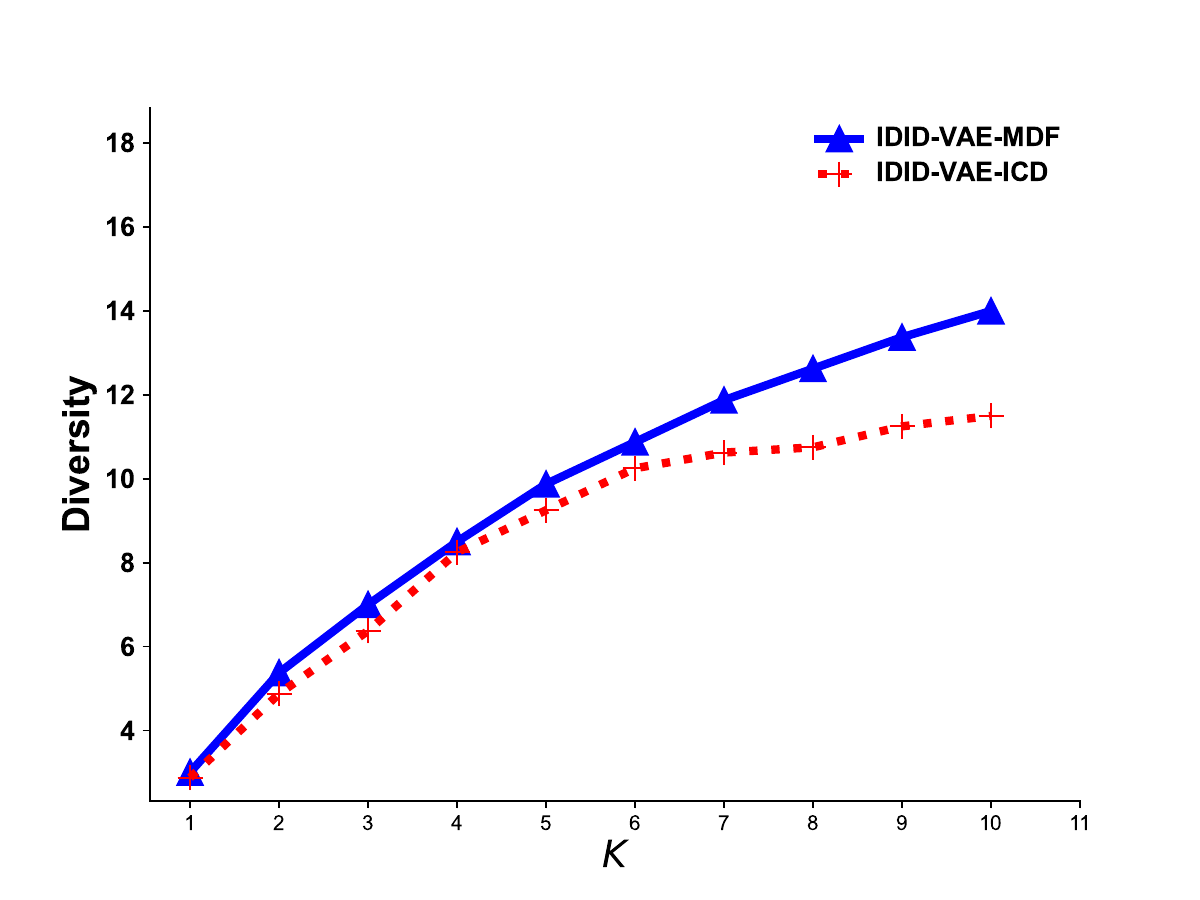}
\label{fig:tiger4_div}}
\caption{For ($a$) $T$ = 3 and ($b$) $T$ = 4, given different $K$ values, the diversity of the top-$K$ policy trees generated by IDID-VAE-MDF and IDID-VAE-ICD respectively.}
\label{fig:tiger_div}
\end{figure}

We first explore how the top-$K$ selection algorithm impacts the diversity of the policy tree set and investigate the relationship between the selection criteria of the policy tree.
To evaluate the diversity of the top-$K$ policy tree set for agent $j$ generated by IDID-VAE-MDF and IDID-VAE-ICD in the Tiger problem, we conducted the experiments on the correlation between the $K$ values and the diversity~(MDF evaluation metric). As shown in the Fig.~\ref{fig:tiger_div}, there is a small diversity gap between the top-$K$ policy trees selected using the ICD index and those selected using the diversity index MDF. This verifies that the policy trees generated by VAE generally have good diversity, which makes it difficult for MDF to distinguish between policy trees and select possibly true behaviors.

\begin{figure}[htbp]
\centering
\subfigure[T = 3]{
\includegraphics[width=0.47\linewidth]{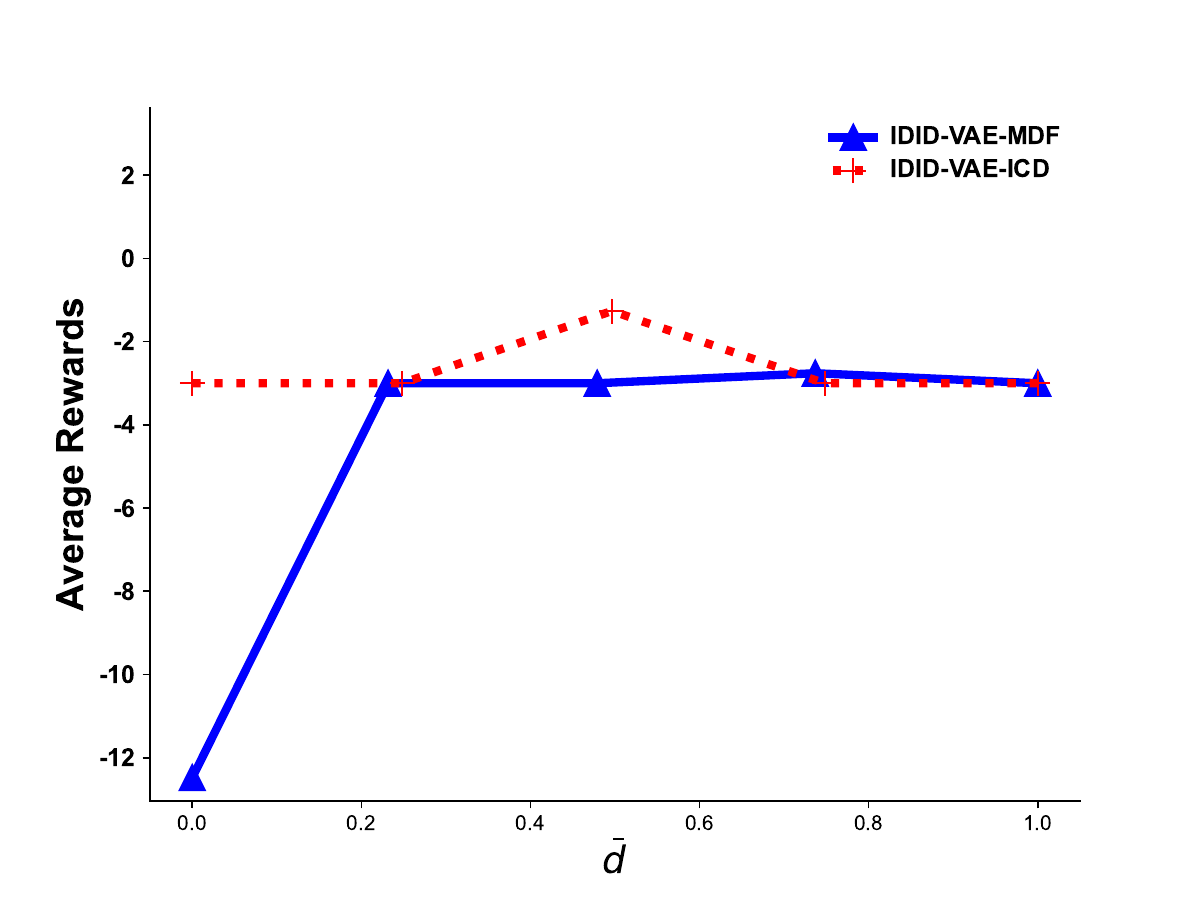}
\label{fig:tiger3_red}}
\hfil
\subfigure[T = 4]{
\includegraphics[width=0.47\linewidth]{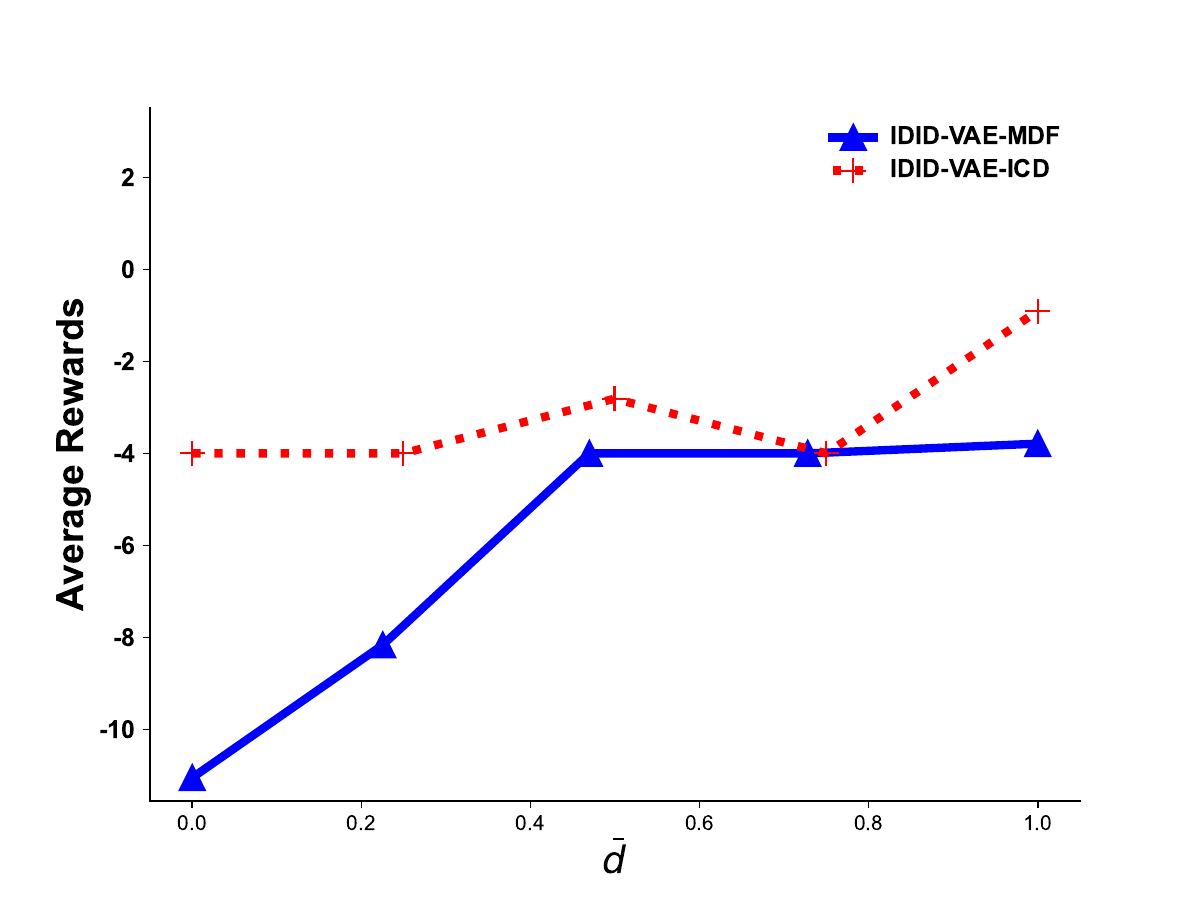}
\label{fig:tiger4_red}}
\caption{For ($a$) $T$ = 3 and ($b$) $T$ = 4, the correlation between the average rewards obtained by agent $i$ and the corresponding metrics~(MDF and ICD) refers to the relationship between the rewards obtained using different models, namely IDID-VAE-MDF and IDID-VAE-ICD.}
\label{fig:tiger_red}
\end{figure}

To explore the correlation between the average reward obtained by the subject agent $i$ and the corresponding metrics~(MDF and ICD), we select the obtained policy tree set using various evaluation metrics to decide the top-$K$ set. The selected metrics are then normalized. Finally, we measure the average reward that agent $i$ can achieve using this policy tree set. As shown in Fig.~\ref{fig:tiger_red}, the rewards obtained by agent $i$ increase as the corresponding metrics rise. The $x$-axis with the label $\bar{d}$ represents the normalized values of ICD and MDF metrics, processed using min-max normalization. Under the policy trees selected by MDF, the average rewards of agent $i$ increase with MDF, but there seems to be an upper limit. The policy trees selected by ICD appear to be less stable. At time slice $T$ = 4, there is an overall upward trend.

\subsubsection{Comparison Results of Multiple I-DID Algorithms}
We investigate the quality of the model, specifically the average rewards, and compare multiple I-DID algorithms. To optimize its decision-making, agent $i$ must anticipate agent $j$'s actions concurrently. We construct an I-DID model tailored for agent $i$ and devise the model space $M_j$ for agent $j$ through multiple I-DID algorithms.
\begin{figure}[htbp]
\centering
\subfigure[T = 3]{
\includegraphics[width=0.45\linewidth]{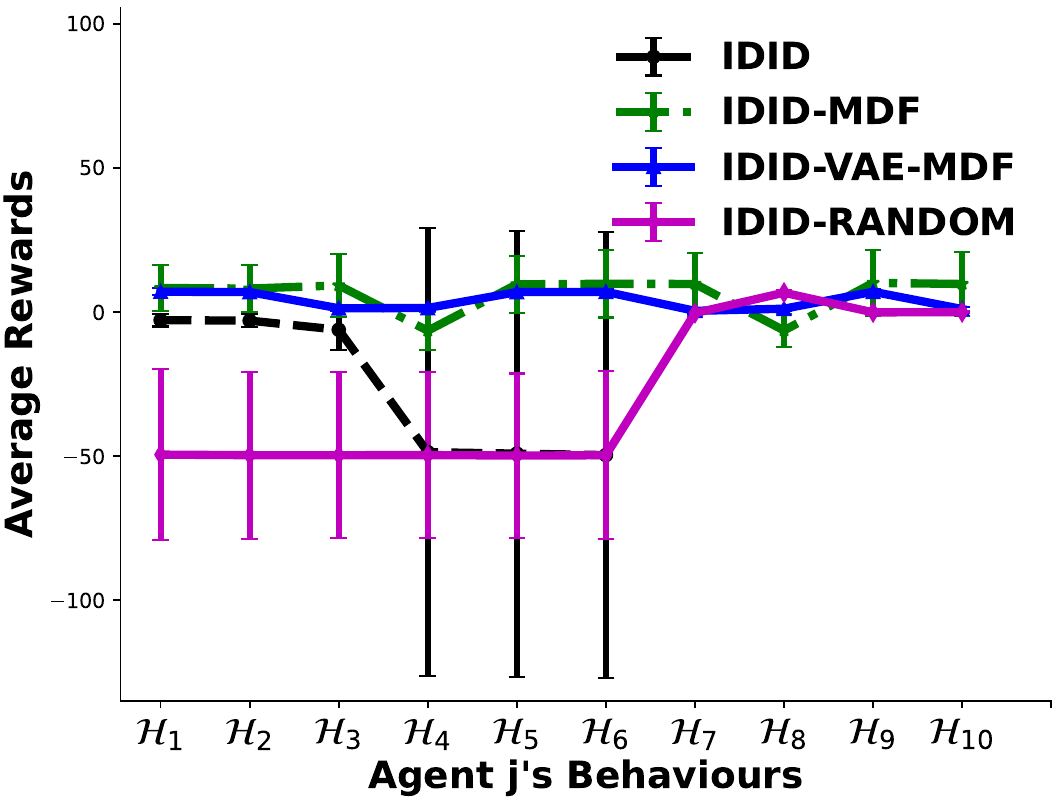}
\label{fig:tiger3_ori}}
\hfil
\subfigure[T = 4]{
\includegraphics[width=0.45\linewidth]{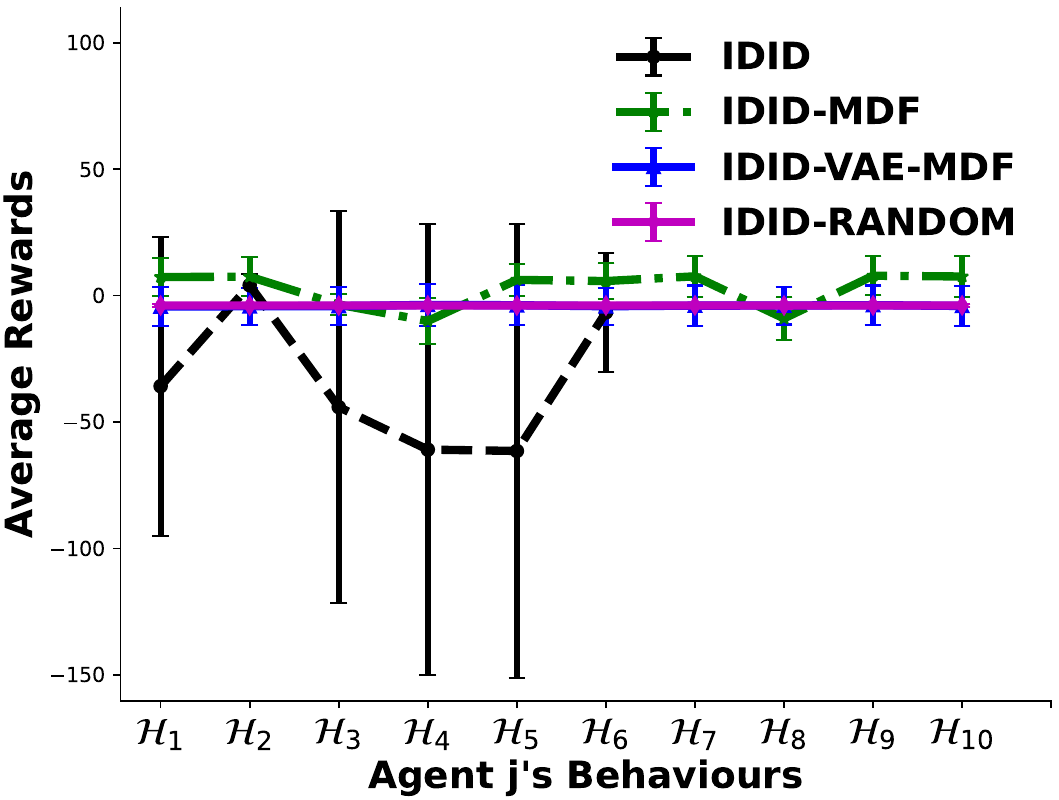}
\label{fig:tiger4_ori}}
\caption{The average rewards are received by the subject agent $i$ with agent $j$'s behaviors models~(generated by IDID, IDID-MDF, IDID-VAE-MDF and IDID-Random) for ($a$) $T$ = 3 and~($b$) $T$ = 4.}
\label{fig:tiger_ori}
\end{figure}
\begin{figure}[htbp]
\centering
\subfigure[T = 3]{
\includegraphics[width=0.45\linewidth]{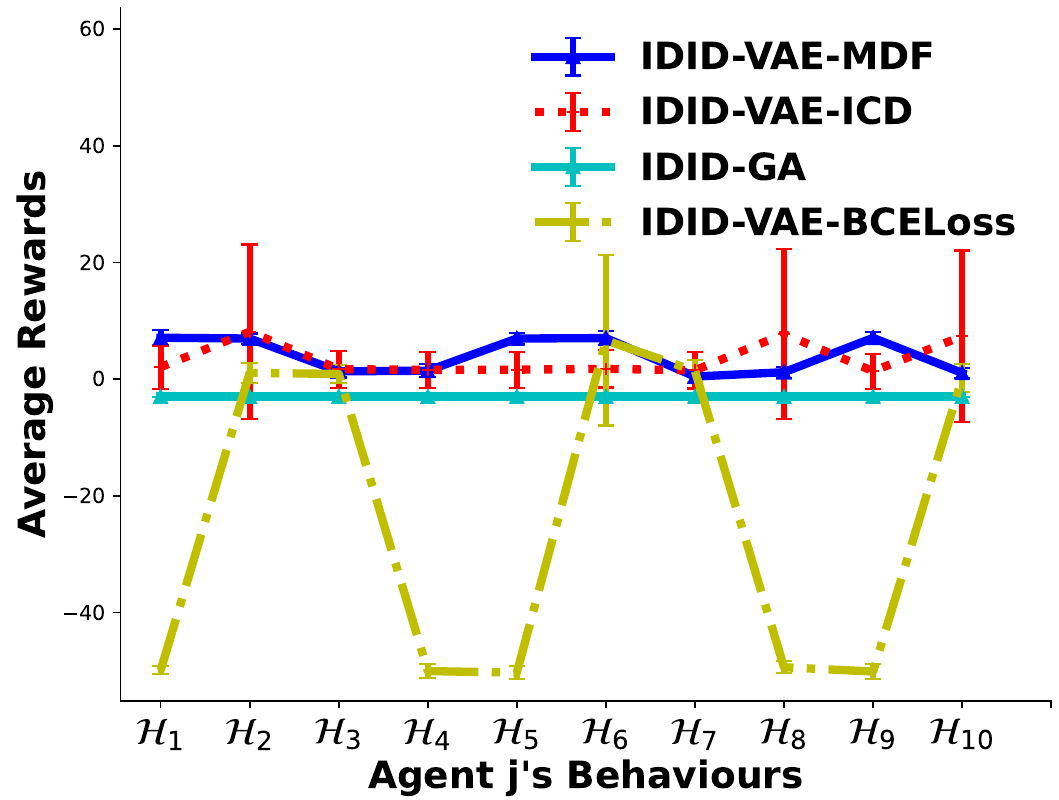}
\label{fig:tiger3_new}}
\subfigure[T = 4]{
\includegraphics[width=0.45\linewidth]{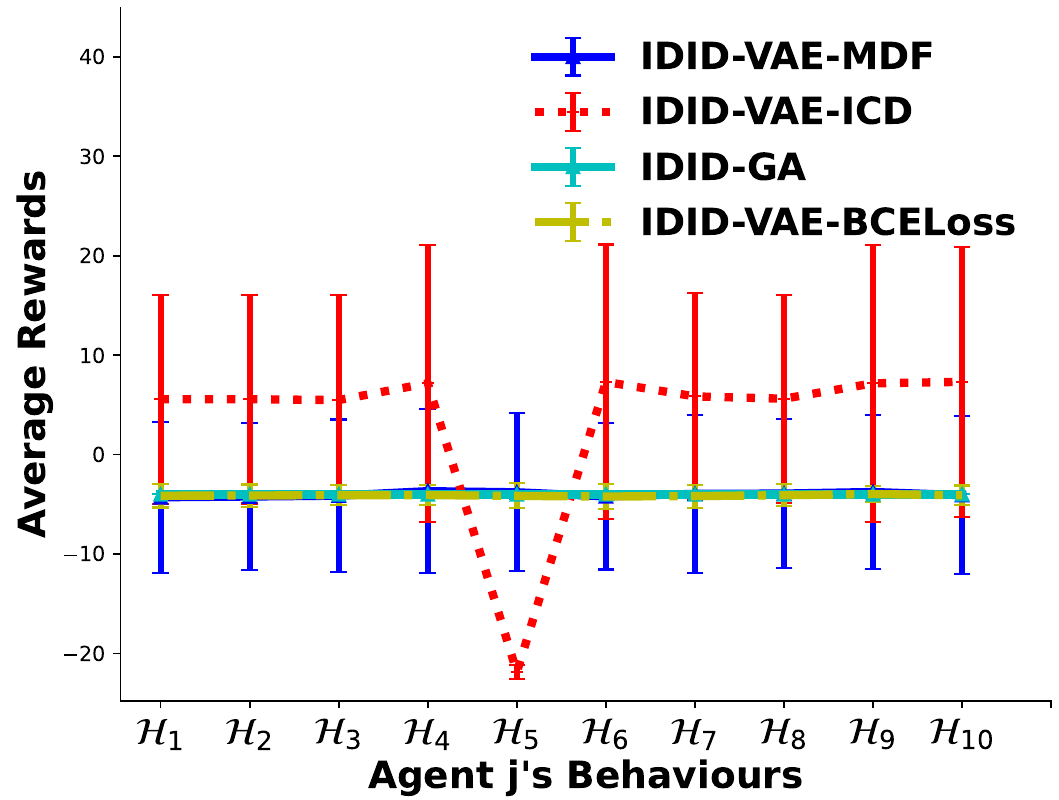}
\label{fig:tiger4_new}}
\caption{The average rewards are received by the subject agent $i$ with agent $j$'s behaviors models~(generated by IDID-VAE-MDF, IDID-VAE-ICD, IDID-GA and IDID-VAE-BCELoss) for ($a$) $T$ = 3 and~($b$) $T$ = 4.}
\label{fig:tiger_new}
\end{figure}

In Figs.~\ref{fig:tiger_ori} and ~\ref{fig:tiger_new}, the results show the average rewards for agent $i$ using various I-DID methods for $T$ = 3 and $T$ = 4. The original I-DID gives agent $j$ six historical models ($M$ = 6). We compared IDID-MDF, IDID-Random, IDID-VAE-MDF, IDID-GA, IDID-VAE-ICD, and IDID-VAE-BCELoss, picking the top 10 models for agent $j$.
We find that IDID-MDF, IDID-VAE-MDF, IDID-GA, and IDID-VAE-ICD all beat the original I-DID. Notably, IDID-VAE-ICD and IDID-MDF do similarly well, probably because the tiger problem isn't too complex. These improved methods are better at modeling and predicting agent $j$'s true behavior, aiding agent $i$'s decisions. In the tests, IDID-VAE-ICD outperformed IDID-VAE-ICD-BCELoss. This is because IDID-VAE-ICD uses a specialized loss based on policy tree nodes' importance, whereas IDID-VAE-ICD-BCELoss relies on the standard binary cross-entropy loss. 
In addition, the VAE-based I-DID solutions outperform the IDID-GA and show their potential in generating agent $j$'s true behaviors from the known models.

\subsection{Multi-agent UAV problems}

\begin{figure}[htbp]
    \centering
    \includegraphics[width=0.225\textwidth]{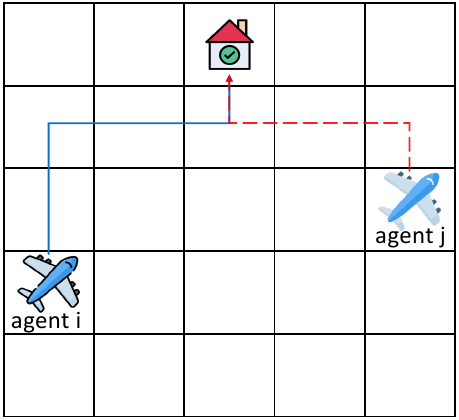}
    \caption{In a two-agent UAV problem, agent $i$ aims to capture agent $j$ before agent $j$ reaches the safe house. The problem specification follows: $|S|=81$, $|A_i|=|A_j|=5$, and $|\Omega_i|=|\Omega_j|=4$.}
    \label{fig:uav}
\end{figure}

The multi-agent Unmanned Aerial Vehicle~(UAV) problem poses a significant challenge in the realm of multi-agent planning. As depicted in Fig.~\ref{fig:uav}, both UAVs, referred to as agents, have the option to move in four different directions or remain stationary. 
In line with realistic scenarios, the UAVs are unable to ascertain the precise positions of other agents and can only receive signals relative to each other. 
Here, we designate agent $i$ as the chaser, tasked with intercepting the fleeing agent $j$, while agent $j$ aims to reach a safe house. 
Since both agents operate concurrently, agent $i$ requires an accurate estimation of agent $j$'s behavior in order to successfully achieve its goal. 
Agent $i$ will be rewarded if it successfully intercepts agent $j$ before it reaches the safe house. We let chaser $i$ use an I-DID model, providing it with potential true behavior models of agent $j$.

\subsubsection{Diversity and Measurements}
\begin{figure}[htbp]
\centering
\subfigure[T = 3]{
\includegraphics[width=0.4725\linewidth]{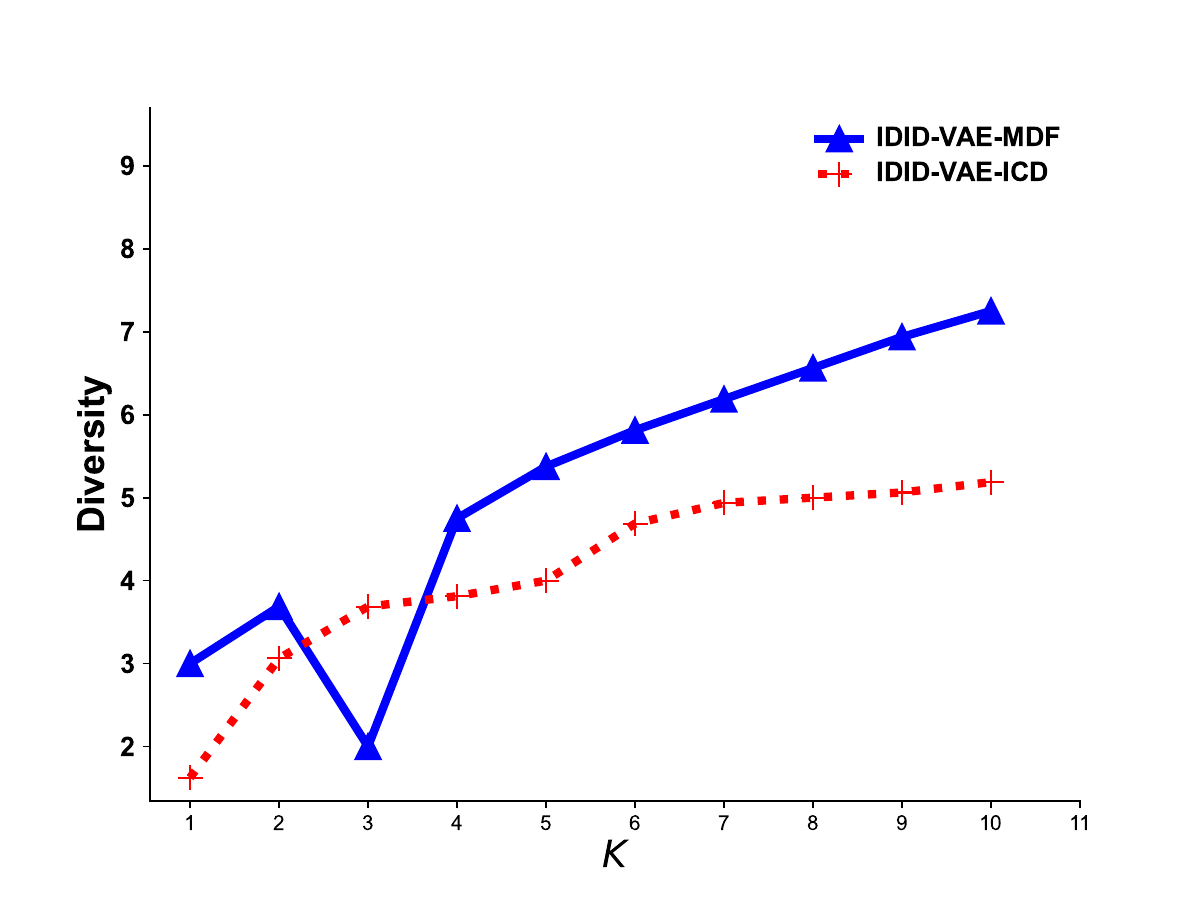}
\label{fig:UAV3_div}}
\hfil
\subfigure[T = 4]{
\includegraphics[width=0.4725\linewidth]{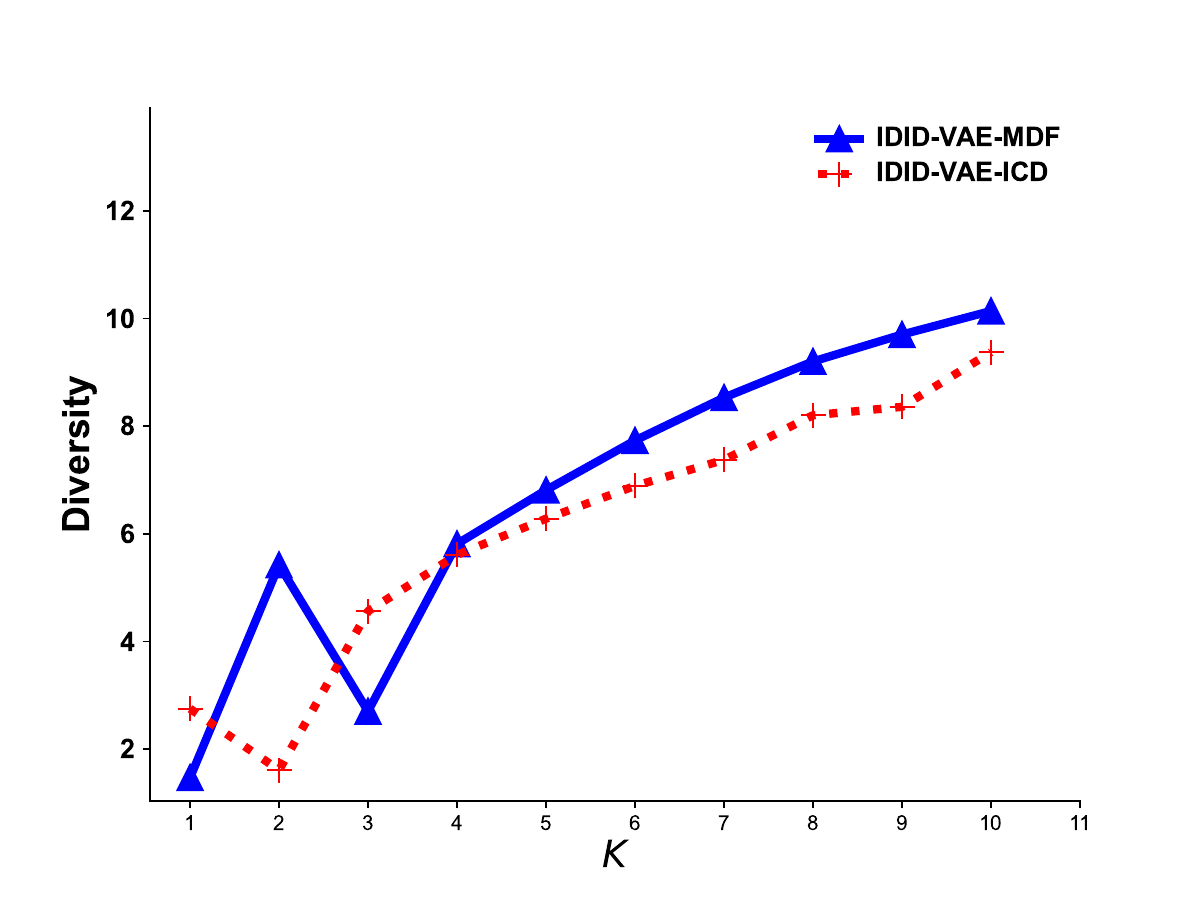}
\label{fig:UAV4_div}}
\caption{For ($a$) $T$ = 3 and ($b$) $T$ = 4, given different $K$ values, the diversity of the top-$K$ policy trees generated by IDID-VAE-MDF and IDID-VAE-ICD respectively. }
\label{fig:UAV_div}
\end{figure}

Initially, we explore how the top-$K$ selection algorithm affects the diversity of the policy tree set, delving into the correlation between policy tree selection criteria.
As illustrated in Fig.~\ref{fig:UAV_div}, both curves display notable fluctuations at small values of $K$, but they gradually rise and stabilize as $K$ increases, ultimately resulting in a slight divergence. By examining the top-$K$ selection function alongside experimental data from two distinct problem domains, it becomes evident that a small $K$ value corresponds to a wide range of potential policy tree set combinations, thereby augmenting the diversity within these sets. This effect may not be immediately obvious in less complex problem domains. Nonetheless, as the complexity of the domain increases, so does the potential decision tree space. Hence, when selecting only a few policy trees from this expanded space, there is a substantial chance of encountering significant disparities among them.
According to the diversity function, smaller $K$ values are associated with a limited number of sets, restricting the potential policy tree paths and sub-trees that can be uncovered. This limitation suggests that the diversity value linked to smaller $K$ will not surpass that of larger $K$ values. As $K$ increases, the final convergence value of IDID-VAE-ICD is lower than that of IDID-VAE-MDF, indicating its superiority in facilitating the selection of an appropriate policy tree set. In essence, continuously augmenting the number of potential policy trees does not inherently lead to greater diversity, nor does it ensure a higher likelihood of discovering the true behavior of agent $j$.

\begin{figure}[htbp]
\centering
\subfigure[T = 3]{
\includegraphics[width=0.47\linewidth]{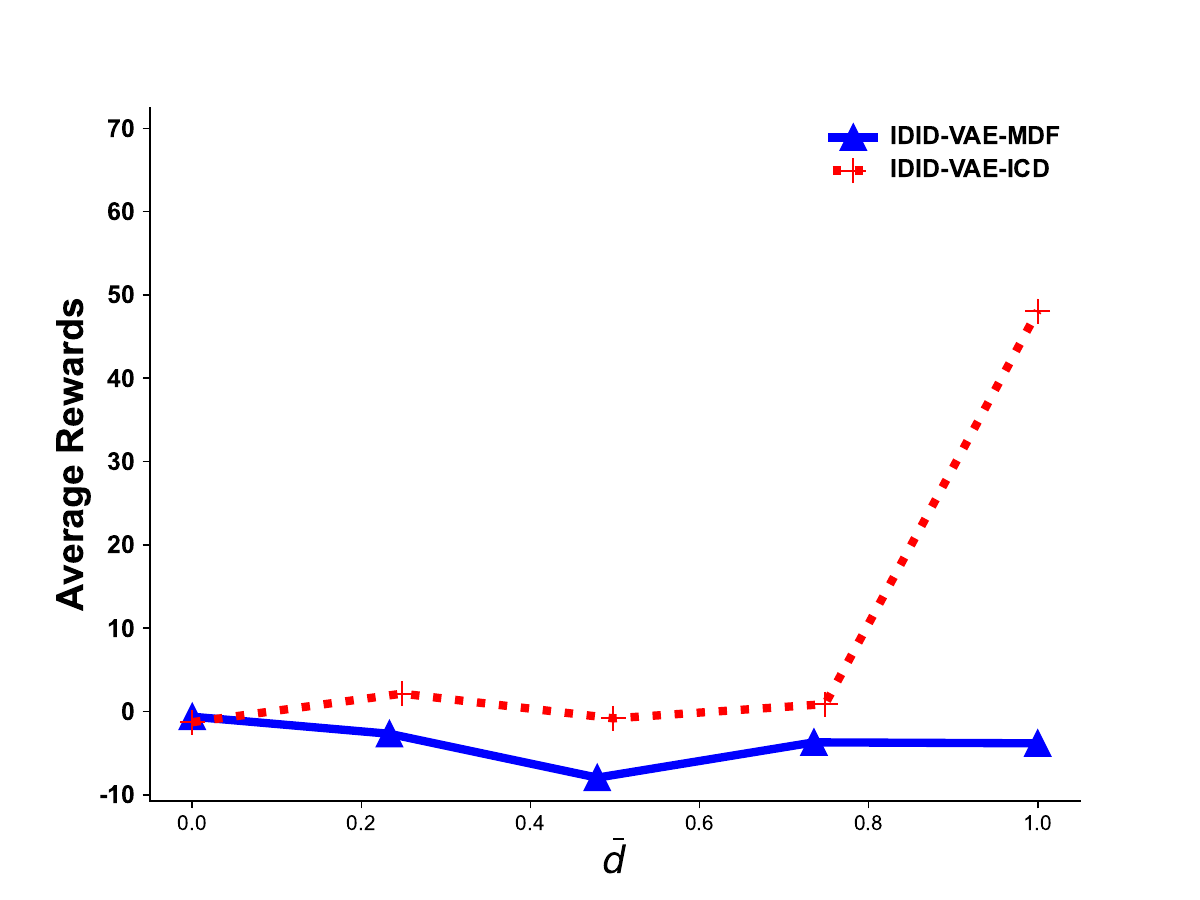}
\label{fig:uav3_red}}
\hfil
\subfigure[T = 4]{
\includegraphics[width=0.47\linewidth]{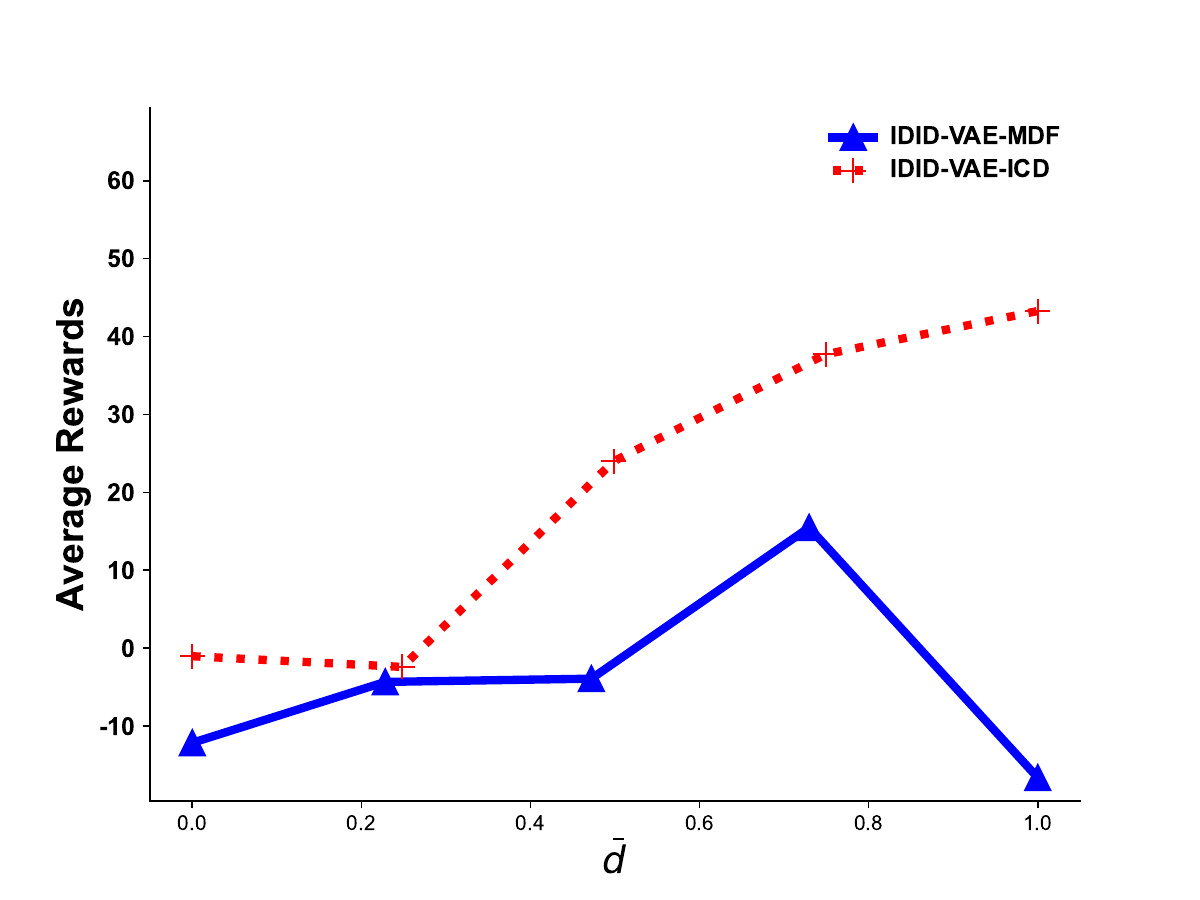}
\label{fig:uav4_red}}
\caption{For ($a$) $T$ = 3 and ($b$) $T$ = 4, the correlation between the average rewards obtained by agent $i$ and the corresponding metrics~(MDF and ICD) refers to the relationship between the rewards obtained using different models, namely IDID-VAE-MDF and IDID-VAE-ICD.}
\label{fig:uav_red}
\end{figure}

We study the correlation between agent $i$'s average reward and two metrics: MDF and ICD. After carefully selecting the top-$K$ policy trees from the available set using various metrics and normalizing them via min-max normalization ($\bar{d}$), we discover a direct correlation between these metrics and agent $i$'s average reward, as illustrated in Fig.~\ref{fig:uav_red}. 
In the Tiger problem (refer to Fig.~\ref{fig:tiger_red}), while the ICD metric does not strongly correlate with the average reward compared to the MDF metric, both still ensure respectable average rewards for the chosen policy trees. However, in the UAV problem, the ICD metric demonstrates a stronger correlation with average rewards than the MDF metric. This suggests that the MDF metric may not consistently identify the true policy tree, particularly given the variety of trees generated by VAE. In contrast, the ICD metric ensures both diversity and a closer alignment with the actual behavior model. This reinforces our observation that VAE-generated policy trees exhibit diversity, posing a challenge for MDF in recognizing and selecting potential real behavior models. When presented with a broad array of policy trees, the MDF metric finds it difficult to discern which ones closely align with the true distribution of agent $j$'s behaviors.

\subsubsection{Comparison Results of Multiple I-DID Algorithms}
We assess model performance by comparing the average rewards across multiple models. For agent $i$ to improve its decisions, it must predict agent $j$'s actions. We are tailoring an I-DID model for agent $i$ and building a model space $M_j$ for agent $j$ using various I-DID techniques. The objective is to measure the accuracy of these methods in modeling and predicting agent $j$'s behavior, thereby influencing agent $i$'s target interception success.
\begin{figure}[htbp]
\centering
\subfigure[T = 3]{
\includegraphics[width=0.45\linewidth]{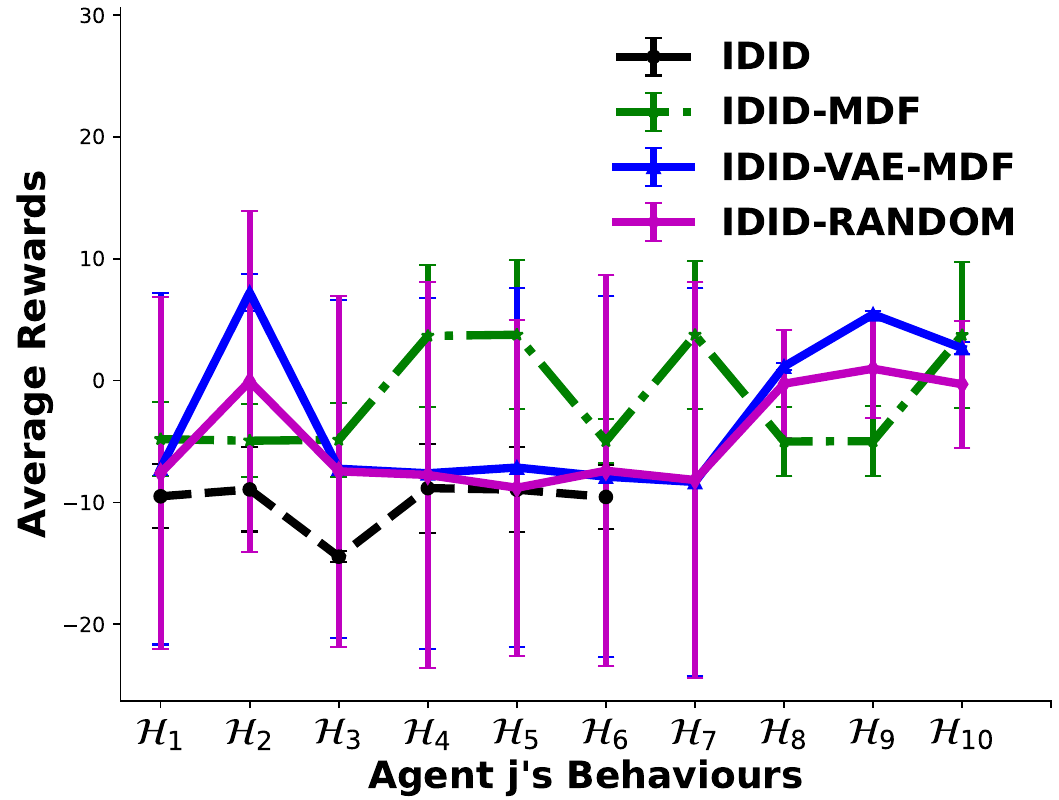}
\label{fig:uav3_ori}}
\subfigure[T = 4]{
\includegraphics[width=0.45\linewidth]{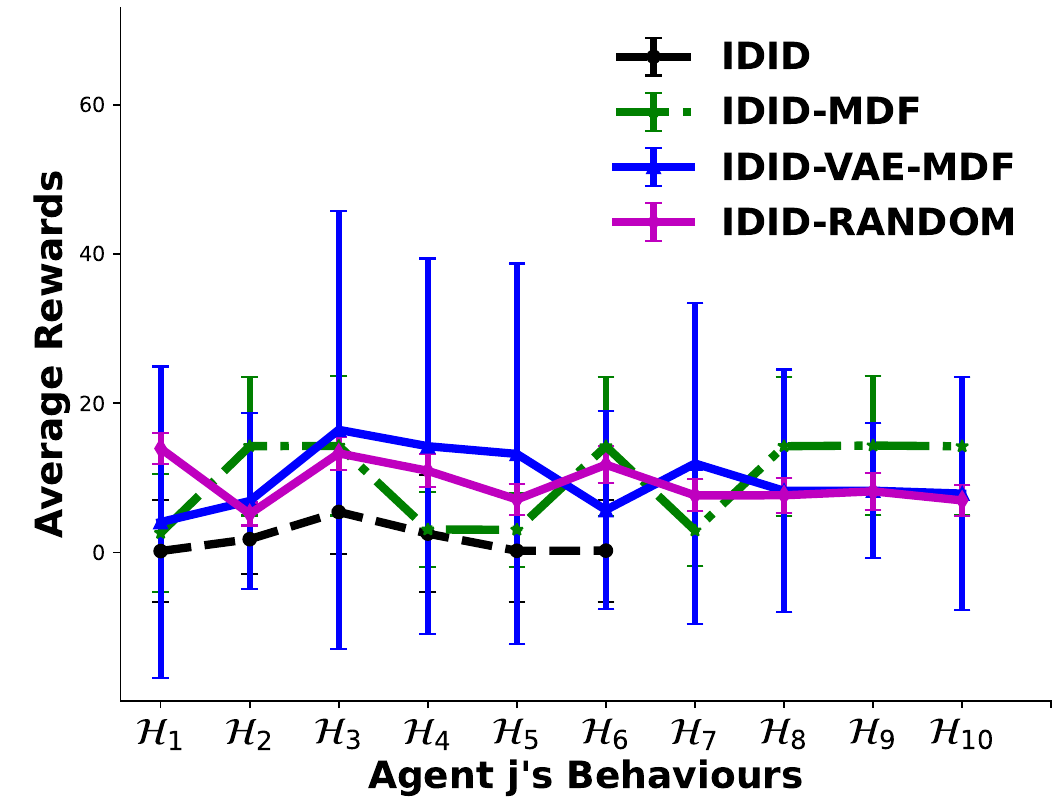}
\label{fig:uav4_ori}}
\caption{The average rewards are received by the subject agent $i$ with agent $j$'s behaviors models~(generated by IDID, IDID-MDF, IDID-VAE-MDF and IDID-Random) for ($a$) $T$ = 3 and~($b$) $T$ = 4.}
\label{fig:uav_ori}
\end{figure}
\begin{figure}[htbp]
\centering 
\subfigure[T = 3]{
\includegraphics[width=0.45\linewidth]{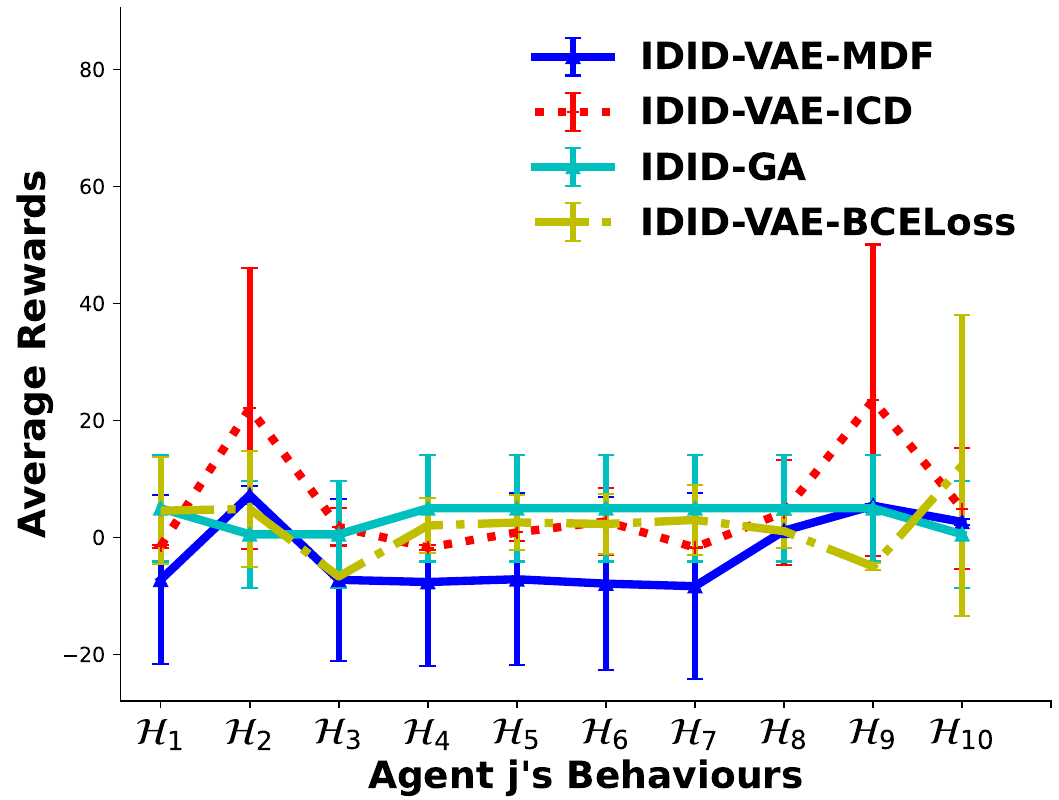}
\label{fig:uav3_new}}
\subfigure[T = 4]{
\includegraphics[width=0.45\linewidth]{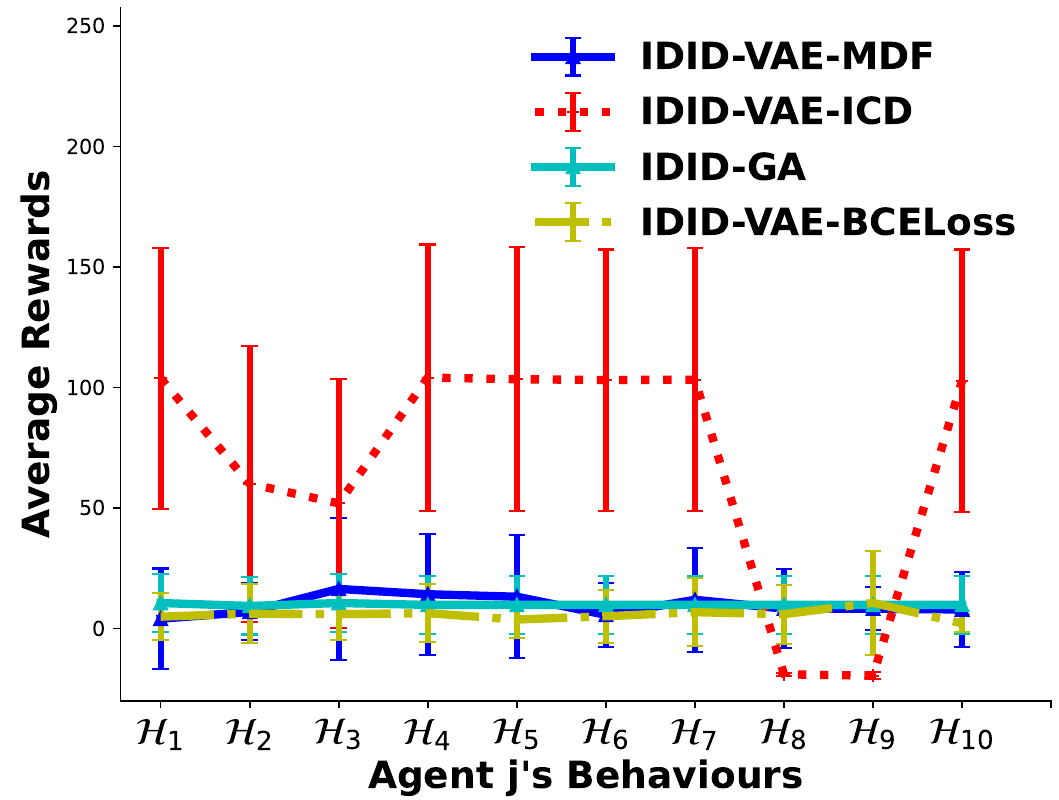}
\label{fig:uav4_new}}
\caption{The average rewards are received by the subject agent $i$ with agent $j$'s behaviors models~(generated by IDID-VAE-MDF, IDID-VAE-ICD, IDID-GA and IDID-VAE-BCELoss) for ($a$) $T$ = 3 and~($b$) $T$ = 4.}
\label{fig:uav_new}
\end{figure}

Figs.~\ref{fig:uav_ori} and ~\ref{fig:uav_new} compare the average rewards in a multi-agent UAV setting. Among the various methods tested, including multiple IDID techniques, IDID-VAE-ICD stands out, outperforming the original I-DID and other approaches. In Figs.~\ref{fig:uav_ori}~(T = 3) and ~\ref{fig:uav_new}~(T = 3), IDID-G, IDID-VAE-BCELoss, and IDID-VAE-ICD surpass other models due to their capability to generate diverse policy trees, comprehensively covering the true model. Notably, in Figs.~\ref{fig:uav_ori}~(T = 4) and ~\ref{fig:uav_new}~(T = 4), unlike IDID-GA and IDID-VAE-BCELoss, IDID-VAE-ICD shines in policy tree generation and selection. This success is partly attributed to the loss function we proposed, which assigns greater weight to the early nodes of the policy tree in long-term decision-making, and partly attributed to the ICD, the construction index of the policy tree set. The VAE model, combined with the ICD index and our policy tree loss function, effectively constructs the authentic subspace within the vast policy tree space, demonstrating its ability to accurately generate agent $j$'s true behaviors from historical interactive data.

\subsection{Experimental Summary and Discussions}
After comparing IDID-VAE-ICD and IDID-VAE-MDF, we find that within the VAE framework, evaluating policy tree diversity using MDF is less effective. This is because VAE can generate more novel paths, maintaining high diversity among new trees, which is a feature absent in the previous methods. ICD leverages the information from the VAE-based policy tree generation process, allowing for an intuitive comparison of different policy trees' ability to represent historical behavior characteristics.

Our experiments reveal that the VAE framework offers a linear relationship between tree generation speed and planning range, providing the speed increase over the previous methods, as depicted in Table~\ref{tab:times}. Although IDID-MDF, IDID-VAE-ICD, and IDID-GA take longer than the basic IDID due to the iterations for the new model generation, IDID-VAE-ICD proves more efficient than IDID-GA. However, since all these algorithms operate offline, their efficiency doesn't hinder the overall performance. Additionally, IDID-VAE-ICD outperforms IDID-VAE-ICD-BCELoss thanks to its customized loss function designed for policy tree node importance distribution.

Comprehensive experiments show that both IDID-MDF and IDID-VAE-ICD outperform IDID-GA and IDID in most cases. IDID-VAE-ICD consistently performs well due to the VAE's ability to balance coherence and individuality in behaviors, while IDID-MDF tends to produce monotonic behaviors. Although finding optimal parameter values remains crucial, the VAE-based I-DID solutions have demonstrated potential in modeling unexpected behaviors that classical AI approaches can't achieve. Despite the randomness in the VAE network, experimental results show stable performance in terms of average rewards, supported by our analysis. Overall, IDID-VAE-ICD exhibits strong adaptability to complex behavioral patterns, maintaining stable performance in uncertain environments, paving the way for future intelligent decision-making systems.
\begin{table}[htbp]
\caption{Running times$\sim$(sec) for all the four methods in the two problem domains generating, specifically focusing on generating $M=10$ policy trees. }
\label{tab:times}
\centering
\begin{tabular}{|c|cc|cc|}
\hline
\textbf{Domain} & \multicolumn{2}{c|}{\textbf{Tiger}} & \multicolumn{2}{c|}{\textbf{UAV}} \\ \hline
\textbf{T} & \multicolumn{1}{c|}{\textbf{3}} & \textbf{4} & \multicolumn{1}{c|}{\textbf{3}} & \textbf{4} \\ \hline
\textbf{IDID} & \multicolumn{1}{c|}{{0.26}} & {32.41} & \multicolumn{1}{c|}{{8.16}} & {406.57} \\ \hline
\textbf{IDID-GA} & \multicolumn{1}{c|}{{8.8}} & {45.1} & \multicolumn{1}{c|}{{27}} & {637.18} \\ \hline
\textbf{IDID-MDF} & \multicolumn{1}{c|}{{442.94}} & {573.85} & \multicolumn{1}{c|}{{81.96}} & {97.40} \\ \hline
\textbf{IDID-VAE-ICD} & \multicolumn{1}{c|}{{2.09}} & {5.36} & \multicolumn{1}{c|}{{9.23}} & {41.84} \\ \hline
\end{tabular}
\end{table}

\section{Conclusion}
\label{sec:final}
We explored neural computing-based I-DID methods to address unpredictable agent behaviors in planning research. Using a VAE approach, we overcame the challenge of reusing incomplete policy trees from interactive data. The VAE model not only generates numerous new policy trees, both complete and incomplete, from a limited set of initial examples, but also offers flexibility in producing varying degrees of deviation. This allows for the creation of a broader range of policy trees.
Furthermore, the integration of a novel perplexity-based metric within our VAE-based approach has significantly enhanced the diversity of the overall policy tree ensemble. By maximizing the coverage of agent $j$'s true policy models, the new method demonstrates promising potential for accurately approximating real agent behavior models from historical data. Following the VAE training with a selection of incomplete policy trees, this data-driven approach has the capability to be applied to increasingly diverse and complex domains, highlighting its adaptability and robustness. 

This work represents a significant contribution to the field of agent planning research, providing a valuable framework for handling the unpredictable behaviors of other agents in multi-agent systems. It also opens the door to exploit neural networks based approaches for addressing the I-DID challenge, which has been addressed through traditional Bayesian approach in the past decade. In the future, we will improve the efficiency, accuracy, and interpret-ability of our models, thereby increasing their adaptability to the uncertainties of online interactive environments. This will be achieved through advancements in deep learning architectures and the incorporation of explainable AI techniques.

\section{Acknowledgments}
This work is supported in part by the National Natural Science Foundation of China~(Grants No.62176225, 62276168 and 61836005) and the Natural  Science Foundation of Fujian Province, China(Grant No. 2022J05176) and Guangdong Province, China(Grant No. 2023A1515010869).

\section*{Appendix}
\label{Appendix}
\noindent {\bf A: [The Four Operators in Reconstructing an Incomplete Policy Tree]}
We present the pseudo-code to implement the four operators in reconstructing an incomplete policy tree. The operators are described in Section~\ref{subsection:RIPT} and in Alg.~\ref{alg:alg2}. As depicted in  Alg.~\ref{alg:OOP},  the key steps of each operator are as follows:
\begin{itemize}
\item {\bf Split Operator:} The {\it split} operator $\mathcal{S}$ divides a sequence into multiple sub-sequences based on a fixed length, and stores them along with their probabilities in the form of a set. 
    \begin{itemize}
    \item Initially, an empty set $H^T$ is initialized to store policy paths and their associated probabilities~(line 4).
    \item The given action-observation sequence is then segmented into multiple sub-sequences of fixed length~(line 5).
    \item These sub-sequences, along with their probabilities, are stored in the set $H^T$~(lines 6-9), providing a structured representation of policy paths.
    \end{itemize}

\item {\bf Union Operator:} The {\it union} operator $\mathcal{U}$ combines the elements of the given set of policy paths into a set of sets of policy trees based on the grouping of their root node actions and observation pairs.
    \begin{itemize}
    \item Policy path subsets $H_{a\textbf{o}}^T$ are defined and their probabilities computed, where each subset begins with a specific action $a$ and observation sequence $\textbf{o}$~(lines 15-16).
    \item For every possible observation sequence $\textbf{o}$ that follows action $a$, the corresponding subsets $H_{a\textbf{o}}^T$ are collected~(lines 14-16).
    \item Sets $\rm{H}_a^T$ are then constructed, encompassing all policy paths that start with a particular action $a$~(line 17).
    \item The overall probability $\mathbb{P}(\rm{H}_a^T)$ of each $\rm{H}_a^T$ is determined by summing the probabilities of its constituent subsets $H_{a\textbf{o}}^T$~(line 18).
    \item Combining the sets $\rm{H}_a^T$ and their probabilities for all actions $a \in A$ yields the comprehensive set of policy paths $\rm{H}^T$~(lines 13-19).
    \item The comprehensive set $\rm{H}^T$ represents all potential policy trees of depth $T$ that can be derived from the original interaction sequence $h^L$~(line 20).
    \end{itemize}

\item {\bf Roulette Operator:} The {\it roulette} operator $\mathcal{R}$ randomly selects an element from a given set.
    \begin{itemize}
    \item Given a set $A$ of elements $(a, p)$ where $p$ is the probability associated with $a$, a random number $p_r$ between 0 and 1 is generated~(line 24).
    \item A running sum $p_c$ is maintained to track the cumulative probabilities encountered while iterating through the set $A$~(line 25).
    \item As soon as $p_c$ exceeds or equals $p_r$, the corresponding element $a$ is returned, effectively selecting an element based on its associated probability~(lines 26-31).
    \end{itemize}

\item {\bf Graphing Operator:}The {\it graphing} operator $\mathcal{G}$ converts the set of paths from the policy tree into a tree structure within a graph model.
    \begin{itemize}
    \item An edge set $E$ is constructed to store the sub-paths within the policy tree~(lines 35-41).
    \item Duplicates are removed from $E$, ensuring that edges with the same time slice, node value~(action), and edge weight~(observation) are not represented multiple times~(line 42).
    \item Unique nodes from the deduplicated edge set $E$ are then extracted and compiled into a node set $V$~(line 44).
    \item Finally, a directed graph $G(E, V)$ is generated and plotted, representing the policy tree constructed from the edge and node sets~(line 45).
    \end{itemize}
\end{itemize}
\begin{algorithm}[htbp]
\caption{The Four Operators in Reconstructing an Incomplete Policy Tree}
 \label{alg:OOP}
 \DontPrintSemicolon  	 
 \SetKwFunction{FME}{$\mathcal{S}$}~ $\triangleleft$ \emph{split operator}\\
 \label{split}
 \SetKwProg{Fn}{Function}{:}{\KwRet  $H^T$}
 \Fn{\FME{ $h^L,\emph{T}$} ~ $\triangleleft$ \emph{$H^\emph{T} \leftarrow \mathcal{S}_\emph{T}{h^L}$}}{     
$a^1o^1,a^2o^2...a^t, o^t,...a^Lo^L\leftarrow \emph{h}^\emph{L}$\;
$H^\emph{T} \leftarrow \emptyset$\;
$\{h^\emph{T}\} \leftarrow \{a^{(l-1)\emph{T}+1}o^{(l-1)\emph{T}+1},\dots,a^{lT}o^{l\emph{T}}\}_{l =1}^{\lfloor \frac{L}{\emph{T}} \rfloor}$\;
\For {$h^\emph{T} \in \{h^\emph{T}\}$}{
    $\mathbb{P}({h^\emph{T}})\leftarrow \#({h_a^\emph{T}})\emph{T}/{L}$\;~ $\triangleleft$ \emph{$\#({h_a^\emph{T}})$ indicates the number of times $h^\emph{T}$ appears in the sequence count the times of $h^L$}
    $H^\emph{T} \leftarrow H^\emph{T} \bigcup (h^\emph{T},\mathbb{P}_{h^\emph{T}})$\;
}
}	
\SetKwFunction{FAE}{$\mathcal{U}$}~ $\triangleleft$ \emph{union operator}\\
 \SetKwProg{Fn}{Function}{:}{\KwRet $\rm{H}^\emph{T}$}
 \Fn{\FAE{ $H^\emph{T}$}~ $\triangleleft$ \emph{$\rm{H}^\emph{T}\leftarrow  \mathcal{U} \emph{H}^\emph{T}$}}{
\For {$a \in A$}{
    \For {$\textbf{o} \in  \Omega^{\emph{T}-1}$}{
        $H_{a\textbf{o}}^\emph{T} \leftarrow \bigcup_{ (h^\emph{T}_{a\textbf{o}},\mathbb{P}({h^\emph{T}_{a\textbf{o}}}))\in H^\emph{T}} (h^\emph{T}_{a\textbf{o}}, \mathbb{P}({h^\emph{T}_{a\textbf{o}}}))$
        $\mathbb{P}({H_{a\textbf{o}}^\emph{T}})\leftarrow \sum_{ (h^\emph{T}_{a\textbf{o}},\mathbb{P}({h^\emph{T}_{a\textbf{o}}}))\in H^\emph{T}} \mathbb{P}({h^\emph{T}_{a\textbf{o}}})$
    }
    $\rm{H}_{a}^\emph{T} \leftarrow \bigcup_{\textbf{o}\in \Omega^{\emph{T}-1}} (\emph{H}_{a\textbf{o}}^\emph{T}, \mathbb{P}({\emph{H}_{a\textbf{o}}^\emph{T}}))$\;
    $\mathbb{P}({\rm{H}_{a}^\emph{T}} )\leftarrow \sum_{\textbf{o}\in \Omega^{\emph{T}-1}} \mathbb{P}({\emph{H}_{a\textbf{o}}^\emph{T}})$
}
$\rm{H}^\emph{T} \leftarrow \bigcup_{a\in A } (\rm H_{a}^\emph{T} ,\mathbb{P}({\rm H_{a}^\emph{T}}))$
}	
\SetKwFunction{FTE}{$\mathcal{R} $}~ $\triangleleft$ \emph{roulette operator}\\ 
\SetKwProg{Fn}{Function}{:}{\KwRet  $a$}
\Fn{\FTE{ $A$} ~ $\triangleleft$ \emph{$a \leftarrow \mathcal{R} A$}}{
    $p_r \leftarrow random()$\;
    $p_c \leftarrow 0$\;
    \For {$(a,p) \in A$}{
        $p_c \leftarrow p_c+p$\;
        \If {$ p_r <= p_c$}{
            \Return$~a$
        }
    }
 }
\SetKwFunction{FAE}{$\mathcal{G}$}~ $\triangleleft$ \emph{Graphing operator}\\
 \SetKwProg{Fn}{Function}{:}{\KwRet  $Tree$}
 \Fn{\FAE{ $\mathcal{H}^T_a$}~ $\triangleleft$ \emph{$Tree \leftarrow\mathcal{G} \mathcal{H}^T_a $}}{
 $E \leftarrow \emptyset$\;
 \For {$(\emph{h}^\emph{T},\mathbb{P}(\emph{h}^\emph{T})) \in \mathcal{H}^T$}{
    $a^1o^1,a^2o^2...a^t, o^t,...a^To^T\leftarrow  \emph{h}^\emph{T}$\;
    \For {$t \in \{1,2,\cdots T-1\}$}{
        $E \leftarrow  E \bigcup (o^t:a^t,a^{t+1})$ 
    }
    }
    $E \leftarrow Deduplicate(E)$\; ~ $\triangleleft$ \emph{deduplicate the edge with same time slice, same node value(action) and edge weight~(observation)}\;
    $V\leftarrow Node(E))$\;
 $Tree \leftarrow G(E,V) $\; 
}	
 \end{algorithm}
\begin{algorithm}[htbp]
\caption{{\it One-Hot} Encoding Operator }
 \label{alg:one-hot}
 \DontPrintSemicolon  	
   \SetKwFunction{FTE}{$\mathcal{I} $}~ $\triangleleft$ \emph{One-hot encoding}\\
\SetKwProg{Fn}{Function}{:}{\KwRet  $\textbf{x}$}
\Fn{\FTE{ $\Tilde{\textbf{x}}$} ~ $\triangleleft$ \emph{$\textbf{x} \leftarrow \mathcal{I}( \Tilde{\textbf{x}})$}}{
$\textbf{x} \in \mathbb{R}^{(|{A}_j|+1)\frac{|\Omega|^T- 1}{|\Omega| -1}}$\;
\For {$l \in \{1,2,\cdots \frac{|\Omega|^T- 1}{|\Omega| -1}\}$}{
    $\textbf{v} \leftarrow \Tilde{\textbf{x}}[(l-1)(|{A}_j|+1)+1:(|{A}_j|+1)l]$\\
    $k \leftarrow \argmax_{k\in \{1,2,\cdots |{A}_j|+1\}} \textbf{v}[k]$\\
    $\textbf{x}[(l-1)(|{A}_j|+1)+1:(|{A}_j|+1)l]\leftarrow  \Tilde{A}_j[k]$
} 
}
\end{algorithm}
\noindent {\bf B: [The Operators in encoding and decoding an Incomplete Policy Tree]}
The ZZOH encoder and decoder operators are described in Section~\ref{subsection:ZZOH} and in Alg.~\ref{alg:alg2}. The operators are used to perform a zigzag transformation on the actions of a given policy tree, encoding them into a binary format using one-hot encoding to form a column vector. Conversely, it can also decode a given column vector from one-hot encoding back into actions, subsequently reconstructing the policy tree. 
As depicted in  Alg.~\ref{alg:ZZOH},  the key steps of each operator are as follows:
\begin{itemize}
\item {\bf ZZOH Encoder Operator:}
    \begin{itemize}
    \item Initially, we extract the action sequence from the given policy tree $\mathcal{H}^T$~(lines 3-12). 
    \item Subsequently, for each action in the sequence, we generate a corresponding one-hot binary code representation $p$~(lines 13-16). This one-hot encoding is designed specifically for policy trees, utilizing the Zig-Zag encoding technique.
    \end{itemize}
\item {\bf ZZOH Decoder Operator:}
    \begin{itemize}
    \item Initially, given a one-hot binary code representation $\textbf{x}$, we decode it back into an action sequence $p$~(lines 20-26). The decoding process reverses the Zig-Zag encoding, accurately reconstructing the original action sequence.
    \item Subsequently, using the decoded action sequence $p$, we reconstruct the policy tree $\mathcal{H}^T$~(lines 27-40). This reconstruction ensures that the generated tree closely matches the original tree, capturing its structure and behavior.
    \end{itemize}
\end{itemize}
\begin{algorithm}[htbp]
\caption{ZZOH Encoder \& Decoder Operators }
 \label{alg:ZZOH}
 \DontPrintSemicolon  	
\SetKwFunction{FSE}{$\mathcal{Z}$}~ $\triangleleft$ \emph{ZZOH encoder operator}\\
 \SetKwProg{Fn}{Function}{:}{\KwRet  $\textbf{x}$}
 \Fn{\FSE{ $\mathcal{H}^T$}~ $\triangleleft$ \emph{$\textbf{x} \leftarrow \mathcal{Z} \mathcal{H}^T$}}{
 ~ $\triangleleft$ \emph{generate action sequence of tree $\mathcal{H}^T$}\;
  $\textbf{x} \in \mathbb{R}^{(|{A}_j|+1)\frac{|\Omega|^T- 1}{|\Omega| -1}}$\;
  $p \leftarrow \{a_l| a_l\leftarrow 0,\forall l \in \{1,2,\cdots \frac{|\Omega|^T- 1}{|\Omega| -1}\}\}$\; 
  \For {$(h^T,i) \in \mathcal{H}^T$}{
  $a^1o^1,a^2o^2...a^t, o^t,...a^To^T\leftarrow  \emph{h}^\emph{T}$\;
 \For {$t \in \{1,2,\cdots T\}$}{
 $l \leftarrow \frac{|\Omega|^{(t-1)}- 1}{|\Omega| -1}+\lfloor \frac{i-1}{|\Omega|^{(t-1)}}\rfloor+1 $\;
     $p[l] \leftarrow a^t$\;
 }
 }
~ $\triangleleft$ \emph{generate one-hot binary code of action sequence $p$}\;
\For {$l \in \{1,2,\cdots \frac{|\Omega|^T- 1}{|\Omega| -1}\}$}{
    $\textbf{x}[(l-1)(|{A}_j|+1)+1:(|{A}_j|+1)l]\leftarrow \Tilde{A}_j^c[p[l]]$
}
}
\SetKwFunction{FSE}{$\mathcal{Z}$}~ $\triangleleft$ \emph{ZZOH decoder operator}\\
 \SetKwProg{Fn}{Function}{:}{\KwRet  $\mathcal{H}^T$}
 \Fn{\FSE{ $\textbf{x}$}~ $\triangleleft$ \emph{$\mathcal{H}^T\leftarrow \mathcal{Z} \textbf{x}$}}{
 ~ $\triangleleft$ \emph{generate action sequence $p$ of one-hot binary code $\textbf{x}$}\;
$p \leftarrow \{a_l| a_l\leftarrow 0,\forall l \in \{1,2,\cdots \frac{|\Omega|^T- 1}{|\Omega| -1}\}\}$\;
\For {$l \in \{1,2,\cdots \frac{|\Omega|^T- 1}{|\Omega| -1}\}$}{
    $\textbf{v} \leftarrow \Tilde{A}_j^c \cdot \textbf{x}[(l-1)(|{A}_j|+1)+1:(|{A}_j|+1)l]$\\
    $k \leftarrow \argmax_{k\in \{1,2,\cdots |{A}_j|+1\}} \textbf{v}[k]$\\
    $p[l] \leftarrow \Tilde{A}_j[k]$
}
~ $\triangleleft$ \emph{generate policy tree from action sequence $p$}\;
$\mathcal{H}^T\leftarrow \bigcup h^T$\;
\For {$(h^T,i) \in \mathcal{H}^T$}{
  $a^1o^1,a^2o^2...a^t, o^t,...a^To^T\leftarrow  \emph{h}^\emph{T}$\;
 \For {$t \in \{1,2,\cdots T\}$}{
 $l \leftarrow \frac{|\Omega|^{(t-1)}- 1}{|\Omega| -1}+\lfloor \frac{i-1}{|\Omega|^{(t-1)}}\rfloor+1 $\;
     $a^t \leftarrow p[l] $\;
     \If{$t\neq T$}{
     $k\leftarrow \lfloor \frac{(i-1)|\Omega|^{t}}{|\Omega|^{(T-1)}}\rfloor+1$\;     
     $o^{t+1} \leftarrow o_{k}$\;
     }
     
 }
 $ h^T \leftarrow a^1o^1,a^2o^2...a^t, o^t,...a^To^T $\;
}
}
\end{algorithm}
The One-hot encoding operator is described in Section~\ref{subsection:VEB} and in Alg.~\ref{alg:alg2}. 
The {\it one-hot} operator $\mathcal{I}$ used to convert the summary vector output by the VAE network into a binary encoding format that matches the one-hot encoding of the policy tree. As depicted in  Alg.~\ref{alg:one-hot},  the key steps of each operator are as follows:  Initialize a vector $\textbf{x}$ to represent the one-hot encoding of the policy tree~(line 3). Then, for each node in the policy tree~(lines 4-8).Locate the index with the maximum value in the corresponding sub-binary representation~(line 6). Represent that sub-binary encoding using the enlarged action space $\Tilde{A}_j$~(line 7).

\clearpage
\bibliographystyle{unsrt}  

\begin{thebibliography}{99}  
  
\bibitem{ref_1_Pan2022}  
Pan, Yinghui, Zhang, Hanyi, Zeng, Yifeng, Ma, Biyang, Tang, Jing, and Ming, Zhong.  
"Diversifying Agent's Behaviors in Interactive Decision Models."  
\emph{Int. J. Intell. Syst.} 37, no. 12 (2022): 12035-12056.  
  
\bibitem{rossijcai15}  
Ross Conroy, Yifeng Zeng, Marc Cavazza, and Yingke Chen.  
"Learning Behaviors in Agents Systems with Interactive Dynamic Influence Diagrams."  
In \emph{IJCAI}, 2015, pp. 39--45.  
  
\bibitem{ref_2_Wang2023}  
Wang, Xindi, Liu, Hao, and Gao, Qing.  
"Data-Driven Decision Making and Near-Optimal Path Planning for Multiagent System in Games."  
\emph{IEEE-J-MASS} 4, no. 3 (2023): 320-328.  
  
\bibitem{ref_3_Buijsse2023}  
Ronald Buijsse, Martijn Willemsen, and Chris Snijders.  
"Data-Driven Decision-Making."  
In \emph{Data Science for Entrepreneurship: Principles and Methods for Data Engineering, Analytics, Entrepreneurship, and the Society}.  
Springer International Publishing, Cham, 2023, pp. 239--277.  
  
\bibitem{ref_4_AN2023}  
Li An, Volker Grimm, Yu Bai, Abigail Sullivan, B.L. Turner, Nicolas Malleson, Alison Heppenstall, Christian Vincenot, Derek Robinson, Xinyue Ye, Jianguo Liu, Emilie Lindkvist, and Wenwu Tang.  
"Modeling agent decision and behavior in the light of data science and artificial intelligence."  
\emph{Environ. Modell. Softw.} 166 (2023): 105713.  
  
\bibitem{ref_5_Pan2022}  
Pan, Yinghui, Ma, Biyang, Zeng, Yifeng, Tang, Jing, Zeng, Buxin, and Ming, Zhong.  
"An Evolutionary Framework for Modelling Unknown Behaviours of Other Agents."  
\emph{IEEE Trans. Emerging Top. Comput. Intell.} 7, no. 4 (2023): 1276-1289.  
  
\bibitem{ref_6_Pan2021TowardDS}  
Yinghui Pan, Jing Tang, Biyang Ma, Yi-feng Zeng, and Zhong Ming.  
"Toward data-driven solutions to interactive dynamic influence diagrams."  
\emph{Knowl. Inf. Syst.} 63, no. 9 (2021): 2431-2453.  
  
\bibitem{ref_7_Nascimento}  
Nascimento, Nathalia, Paulo Alencar, and Donald Cowan.  
"Self-Adaptive Large Language Model (LLM)-Based Multiagent Systems."  
In \emph{2023 IEEE ACSOS-C}, 2023, pp. 104-109.  
  
\bibitem{ref_8_J10498880}  
Wason, Ritika, Parul Arora, Devansh Arora, Jasleen Kaur, Sunil Pratap Singh, and M. N. Hoda.  
"Appraising Success of LLM-based Dialogue Agents."  
In \emph{2024 IEEE INDIACom}, 2024, pp. 1570-1573.  
  
\bibitem{ref_9_10520238}  
Lu, Jiaying, Bo Pan, Jieyi Chen, Yingchaojie Feng, Jingyuan Hu, Yuchen Peng, and Wei Chen.  
"AgentLens: Visual Analysis for Agent Behaviors in LLM-based Autonomous Systems."  
\emph{IEEE Trans. Visual Comput. Graphics}, 2024, pp. 1-17.  
  
\bibitem{ref_10_zhang2023explaining}  
Zhang, Xijia, Yue Guo, Simon Stepputtis, Katia Sycara, and Joseph Campbell.  
"Explaining Agent Behavior with Large Language Models."  
\emph{arXiv}, 2023, eprint 2309.10346.  
  
\bibitem{ref_11_BERGER2024106003}  
Berger, Uta, Andrew Bell, C. Michael Barton, Emile Chappin, Gunnar Dreßler, Tatiana Filatova, Thibault Fronville, Allen Lee, Emiel {van Loon}, Iris Lorscheid, Matthias Meyer, Birgit Müller, Cyril Piou, Viktoriia Radchuk, Nicholas Roxburgh, Lennart Schüler, Christian Troost, Nanda Wijermans, Tim G. Williams, Marie-Christin Wimmler, and Volker Grimm.  
"Towards reusable building blocks for agent-based modelling and theory development."  
\emph{Environ. Modell. Softw.}, vol. 175, 2024, p. 106003.  
  
\bibitem{ref_11_9527397}  
Zhang, Wei, Andrea Valencia, and Ni-Bin Chang.  
"Synergistic Integration Between Machine Learning and Agent-Based Modeling: A Multidisciplinary Review."  
\emph{IEEE Trans. Neural Networks Learn. Syst.}, vol. 34, no. 5, 2023, pp. 2170-2190.  
  
\bibitem{ref_12_Hazra2023}  
Hazra, Rishi, and Luc De Raedt.  
"Deep Explainable Relational Reinforcement Learning: A Neuro-Symbolic Approach."  
In \emph{Machine Learning and Knowledge Discovery in Databases: Research Track}, 2023, pp. 213--229.  
Cham: Springer International Publishing.  
  
\bibitem{ref_13_Belle2024}  
Belle, Vaishak, Michael Fisher, Alessandra Russo, Ekaterina Komendantskaya, and Alistair Nottle.  
"Neuro-Symbolic AI + Agent Systems: A First Reflection on Trends, Opportunities and Challenges."  
In \emph{AAMAS Workshops}, 2024, pp. 180--200.  
Cham: Springer Nature Switzerland.  
  
\bibitem{ref_14_Kononov2023}  
Kononov, Roman, and Oleg Maslennikov.  
"Performing decision-making tasks through dynamics of recurrent neural networks trained with reinforcement learning."  
In \emph{Proc. 2023 DCNA}, 2023, pp. 144-147.  
  
\bibitem{ref_15_9904958}  
Wang, Xu, Sen Wang, Xingxing Liang, Dawei Zhao, Jincai Huang, Xin Xu, Bin Dai, and Qiguang Miao.  
"Deep Reinforcement Learning: A Survey."  
\emph{IEEE Trans. Neural Networks Learn. Syst.}, vol. 35, no. 4, 2024, pp. 5064-5078.  
  
\bibitem{ref_16_vae2021}  
Almaeen, Manal, Yasir Alanazi, Nobuo Sato, W. Melnitchouk, Michelle P. Kuchera, and Yaohang Li.  
"Variational Autoencoder Inverse Mapper: An End-to-End Deep Learning Framework for Inverse Problems."  
In \emph{Proc. 2021 IJCNN}, 2021, pp. 1-8.  
  
\bibitem{ref_17_MO-MIX}  
Hu, Tianmeng, Biao Luo, Chunhua Yang, and Tingwen Huang.  
"MO-MIX: Multi-Objective Multi-Agent Cooperative Decision-Making With Deep Reinforcement Learning."  
\emph{IEEE Trans. Pattern Anal. Mach. Intell.}, vol. 45, no. 10, Oct. 2023, pp. 12098–12112.  
  
\bibitem{ref_18_YANG2024109754}  
Sen Yang, Yi Zhang, Xinzheng Lu, Wei Guo, and Huiquan Miao.  
"Multi-agent deep reinforcement learning based decision support model for resilient community post-hazard recovery."  
\emph{Reliab. Eng. Syst. Safe.}, vol. 242, 2024, p. 109754.  
  
\bibitem{ref_19_le2022deep}  
Le, Ngan, Vidhiwar Singh Rathour, Kashu Yamazaki, Khoa Luu, and Marios Savvides.  
"Deep reinforcement learning in computer vision: a comprehensive survey."  
\emph{Artif. Intell. Rev.}, 2022, pp. 1--87.  
  
\bibitem{ref_20_zhu2021deep}  
Zhu, Kai, and Tao Zhang.  
"Deep reinforcement learning based mobile robot navigation: A review."  
\emph{Tsinghua Science and Technology}, vol. 26, no. 5, 2021, pp. 674--691.  
  
\bibitem{ref_21_wang2021scc}  
Wang, Xiangjun, Junxiao Song, Penghui Qi, Peng Peng, Zhenkun Tang, Wei Zhang, Weimin Li, Xiongjun Pi, Jujie He, Chao Gao, and others.  
"SCC: An efficient deep reinforcement learning agent mastering the game of StarCraft II."  
In \emph{Proc. 2021 ICML}, 2021, pp. 10905--10915.  
  
\bibitem{ref_22_Zeng2012}  
Zeng, Yifeng, and Prashant Doshi.  
"Exploiting Model Equivalences for Solving Interactive Dynamic Influence Diagrams."  
\emph{J. Artif. Int. Res.}, vol. 43, no. 1, Jan. 2012, pp. 211–255.  
  
\bibitem{ref_23_Pan2015}  
Pan, Yinghui, Yifeng Zeng, Yanping Xiang, Le Sun, and Xuefeng Chen.  
"Time-Critical Interactive Dynamic Influence Diagram."  
\emph{Int. J. Approx. Reasoning}, vol. 57, no. C, Feb. 2015, pp. 44–63.  
  
\bibitem{ref_24_Pan2015I}  
Pan, Yinghui, Yifeng Zeng, and Hua Mao.  
"Learning Agents' Relations in Interactive Multiagent Dynamic Influence Diagrams."  
In \emph{ADMI 2014. LNCS}, 2015, pp. 1--11.  
Cham: Springer International Publishing.  
  
\bibitem{ref_25_ALBRECHT2018}  
Stefano V. Albrecht and Peter Stone.  
"Autonomous agents modelling other agents: A comprehensive survey and open problems."  
\emph{arXiv}, 2017, vol. abs/1709.08071.  
  
\bibitem{ref_25_10.1007/11527862_33}  
Masoumeh Tabaeh Izadi.  
"Sequential Decision Making Under Uncertainty."  
In \emph{Abstraction, Reformulation and Approximation}, 2005, pp. 360--361.  
Berlin, Heidelberg: Springer Berlin Heidelberg.  
  
\bibitem{ref_26_Zeng2016Approximating}  
Zeng, Yifeng, Prashant Doshi, Yingke Chen, Yinghui Pan, Hua Mao, and Muthukumaran Chandrasekaran.  
"Approximating Behavioral Equivalence for Scaling Solutions of I-DIDs."  
\emph{Knowl. Inf. Syst.}, vol. 49, no. 2, Nov. 2016, pp. 511–552.  
  
\bibitem{ref_27_Hinton06:Reducing}  
Geoffrey E. Hinton and Ruslan R. Salakhutdinov.  
"Reducing the Dimensionality of Data with Neural Networks."  
\emph{Science}, vol. 313, no. 5786, 2006, pp. 504--507.  


\bibitem{ref_28_Kingma2013AutoEncodingVB}  
Diederik P. Kingma and Max Welling.  
"Auto-Encoding Variational Bayes."  
\emph{CoRR}, abs/1312.6114, 2013.  
  
\bibitem{ref_29_Zeng12:Exploiting}  
Zeng, Yifeng, and Prashant Doshi.  
"Exploiting Model Equivalences for Solving Interactive Dynamic Influence Diagrams."  
\emph{J. Artif. Intell. Res.}, vol. 43, 2012, pp. 211-255.  
  
\bibitem{ref_30_DoshiGD20}  
Prashant Doshi, Piotr J. Gmytrasiewicz, and Edmund H. Durfee.  
"Recursively modeling other agents for decision making: A research perspective."  
\emph{Artif. Intell.}, vol. 279, 2020.  
  
\bibitem{ref_31_Doshi09:Graphical}  
Prashant Doshi, Yifeng Zeng, and Qiongyu Chen.  
"Graphical Models for Interactive POMDPs: Representations and Solutions."  
\emph{Auton. Agent Multi-ag.}, vol. 18, no. 3, 2009, pp. 376-416.  
  
\bibitem{ref_32_Gmytrasiewicz05:Framework:JAIR}  
Gmytrasiewicz P. J. and Doshi P.  
"A Framework for Sequential Planning in Multiagent Settings."  
\emph{J. Artif. INntell. Res.}, vol. 24, 2005, pp. 49-79.  
  
\bibitem{ref_33_Seuken07:Formal}  
Seuken, S., and S. Zilberstein.  
"Formal models and algorithms for decentralized decision making under uncertainty."  
\emph{Auton. Agent Multi-ag.}, vol. 17, no. 2, 2008, pp. 190--250.  
  
\bibitem{ref_34_Chen2015}  
Chen, Yingke, Prashant Doshi, and Yifeng Zeng.  
"Iterative Online Planning in Multiagent Settings with Limited Model Spaces and PAC Guarantees."  
In \emph{AAMAS '15}, 2015, pp. 1161–1169.  
Richland, SC: AAAI Press.  
  
\bibitem{ref_35_Ding2021TheRF}  
Ming Ding.  
"The road from MLE to EM to VAE: A brief tutorial."  
\emph{AI Open}, vol. 3, 2021, pp. 29-34.  
  
\bibitem{ref_36_8365805}  
Rizk, Yara, Mariette Awad, and Edward W. Tunstel.  
"Decision Making in Multiagent Systems: A Survey."  
\emph{IEEE Trans. Cognit. Dev. Syst.}, vol. 10, no. 3, 2018, pp. 514-529.  
  
\bibitem{ref_37_9455523}  
Chen, Wenbin, Guo Xie, Wenjiang Ji, Rong Fei, Xinhong Hei, Siyu Li, and Jialin Ma.  
"Decision Making for Overtaking of Unmanned Vehicle Based on Deep Q-learning."  
In \emph{2021 IEEE DDCLS}, 2021, pp. 350-353.  
  
\bibitem{ref_38_10172334}  
Wang, Xindi, Hao Liu, and Qing Gao.  
"Data-Driven Decision Making and Near-Optimal Path Planning for Multiagent System in Games."  
\emph{IEEE J. Miniaturization Air Space Syst.}, vol. 4, no. 3, 2023, pp. 320-328.  
  
\bibitem{ref_39_9119863}  
Sun, Changyin, Wenzhang Liu, and Lu Dong.  
"Reinforcement Learning With Task Decomposition for Cooperative Multiagent Systems."  
\emph{IEEE Trans. Neural Networks Learn. Syst.}, vol. 32, no. 5, 2021, pp. 2054-2065.  
  
\bibitem{ref_40_9512658}  
Habib, Maki K., Samuel A. Ayankoso, and Fusaomi Nagata.  
"Data-Driven Modeling: Concept, Techniques, Challenges and a Case Study."  
In \emph{2021 IEEE ICMA}, 2021, pp. 1000-1007.  
  
\bibitem{ref_41_9675815}  
Abroshan, Mahed, Kai Hou Yip, Cem Tekin, and Mihaela van der Schaar.  
"Conservative Policy Construction Using Variational Autoencoders for Logged Data With Missing Values."  
\emph{IEEE Trans. Neural Networks Learn. Syst.}, vol. 34, no. 9, 2023, pp. 6368-6378.  
  
\bibitem{ref_43_Conroy2015}  
Conroy, Ross, Yifeng Zeng, Marc Cavazza, and Yingke Chen.  
"Learning Behaviors in Agents Systems with Interactive Dynamic Influence Diagrams."  
In \emph{Proc. 2015 IJCAI}, 2015, pp. 39–45.  
Buenos Aires, Argentina: AAAI Press.  
  
\bibitem{ref_42_Zeng2012Improved}  
Zeng, Yifeng, Hua Mao, Yinghui Pan, and Jian Luo.  
"Improved Use of Partial Policies for Identifying Behavioral Equivalence."  
In \emph{Proc. 2012 AAMAS}, vol. 2, 2012, pp. 1015–1022.  
Richland, SC: AAAI Press.  
  
\bibitem{ref_44_Doshi2009}  
Doshi, Prashant, Yifeng Zeng, and Qiongyu Chen.  
"Graphical models for interactive POMDPs: representations and solutions."  
\emph{Autonomous Agents and Multi-Agent Systems}, vol. 18, no. 3, June 2009, pp. 376–416.  
  

\bibitem{ref_45_mc21}  
Zeng, Yifeng, Ran, Qiang, Ma, Biyang, and Pan, Yinghui.  
"Modelling other agents through evolutionary behaviours."  
\emph{Memet. Comput.} 14, no. 1 (2022): 19--30.  
  
\end{thebibliography}

\end{document}